\documentclass[preprint2]{aastex}		

\usepackage{amssymb}
\usepackage{amsthm}
\usepackage{amsmath}
\usepackage{xcolor}
\usepackage{natbib}
\usepackage[normalem]{ulem}
\usepackage[above]{placeins}
\usepackage{hyperref}
\usepackage{array}

\newcommand{\new}[1]{#1}
\newcommand{\old}[1]{}

\newcommand{\stephan}[1]{\textcolor{blue}{{#1}}}

\renewcommand{\vec}[1]{\mathbf{#1}}

\shorttitle{Anelastic versus Compressible}
\shortauthors{Verhoeven et al.}

\begin{document}


\title{ANELASTIC VERSUS FULLY COMPRESSIBLE TURBULENT  \new{RAYLEIGH-B\'ENARD} CONVECTION}


\author{Jan Verhoeven, Thomas Wieseh\"ofer and Stephan Stellmach}
\affil{Institut f\"ur Geophysik, Westf\"alische Wilhelms Universit\"at M\"unster, Germany}

\email{jan.verhoeven@uni-muenster.de}




\begin{abstract}
Numerical simulations of turbulent \new{Rayleigh-B\'enard} convection in an ideal gas, using either the anelastic approximation or the fully compressible equations, are compared. Theoretically, the anelastic approximation is expected to hold in weakly superadiabatic systems with $\epsilon = \Delta T / T_r \ll 1$, where $\Delta T$ denotes the superadiabatic temperature drop over the convective layer and $T_r$ the bottom temperature. 
Using direct numerical simulations\old{ in a plane layer geometry}, a \new{systematic}\old{detailed} comparison of anelastic and fully compressible convection is carried out. With decreasing superadiabaticity $\epsilon$, the fully compressible results are found to converge linearly to the anelastic solution  with larger density contrasts generally improving the match. We conclude that in many solar and planetary applications, where the superadiabaticity is expected to be vanishingly small, results obtained with the anelastic approximation are in fact more accurate than fully compressible computations, which typically fail to reach small $\epsilon$ for numerical reasons. On the other hand, if the astrophysical system studied contains $\epsilon\sim O(1)$ regions, such as the solar photosphere, fully compressible simulations have the advantage of capturing the full physics. Interestingly, even in weakly superadiabatic regions, like the bulk of the solar convection zone, the errors introduced by using artificially large values for $\epsilon$ for efficiency reasons remain moderate. If quantitative errors of the order of $10\%$ are acceptable in such low $\epsilon$ regions, our work suggests that fully compressible simulations can indeed be computationally more efficient than their anelastic counterparts.
\end{abstract}





\section{INTRODUCTION}
\label{intro}

Thermal convection is of primary importance in astrophysical objects. It carries the heat flow over large regions in stellar and planetary interiors, and is one of the major sources of mechanical mixing in these objects. Furthermore, some of the most striking, large-scale features of stars and planets are powered by convective motions, such as intrinsic dynamo-generated magnetic fields, plate tectonics on Earth and possibly also the zonal winds on Jupiter and other giant planets (e.g. \citealp{Brun2004,Brandenburg2005,Trompert1998,Tackley2000,Bercovici2003,Heimpel2005,Verhoeven2014}).

The convective regions in stellar and planetary objects typically feature a non-negligible density stratification, and the flows are often subsonic. In this paper, we compare two approaches that are commonly used for modeling convection in these systems numerically---the fully compressible approach and the so-called anelastic approximation. Our goal is to quantify the accuracy and efficiency of both methods in a given situation, guiding modelers in making the right choice for their particular problem at hand. 

The fully compressible equations are the most fundamental equations governing thermal convection. They can directly be derived from first principles of physics, such as mass, energy, and momentum conservation, equipped with constitutive relations that characterize the fluid. The resulting equations are thus very general and encompass the full range of physical behavior, from the temporal evolution of the convective motions to the propagation of sound waves. On the one hand, this allows to study regions such as the outermost layers of the Sun, where the Mach number, i.e. the ratio of convective velocity to the sound speed, becomes $O(1)$. On the other hand, problems arise in low Mach number regions where the flow velocities are much slower than the sound speed, which is typically the case in the bulk of deep stellar and planetary interiors. Even though the convective motions in such regions occur on time scales which are many orders of magnitude larger than the acoustic time scale, standard numerical schemes have to explicitly resolve the sound waves for stability reasons. This forces modelers to assume artificially large Mach numbers, which reduces the differences between the convective and acoustic time scales to numerically tractable values\new{ (e.g. \citealp{Tobias1998,Brummell2002,Kapyla2010})}. Errors introduced by this procedure occur as an unavoidable side-effect in the fully compressible framework. Still, most of the numerical resources typically go into capturing acoustic wave propagation phenomena, which are generally believed to be irrelevant for the investigated convection dynamics\new{ (but see \citealp{Bogdan1993,Meakin2006})}.

To circumvent the problems arising from the numerical stiffness of the fully compressible equations, different "sound-proof" models, such as the low Mach number approach (e.g. \citealp{Majda1985,Bell2004,Almgren2006}), the pseudo-incompressible approximation \citep{Durran1989} or the anelastic approximation \citep{Batchelor1953,Ogura1962,Gough1969,Gilman1981,Lantz1999}
have been developed.  Instead of prescribing artificially {\em large} Mach numbers, all these approaches take the opposite route by considering the {\em small} Mach number limit of the fully compressible equations. The same time scale disparities which make solving the fully compressible equations numerically challenging are thus exploited to substantially simplify the equations. As a result, the pressure field adapts instantaneously, which effectively filters out the sound waves. This comes, however, at the price of loosing the ability to study regions where the Mach number is not small.

Among the sound-proof approaches described above, the anelastic approximation is the one most commonly deployed for modeling stellar and planetary interiors (e.g. \citealp{Glatzmaier1996a,Miesch2008,Brun2011,Jones2011}).
The anelastic equations are theoretically predicted to hold for low Mach number systems in which only slight thermodynamic perturbations from a hydrostatic background state occur (e.g. \citealp{Gough1969}). In convective systems, the background state is typically assumed to be adiabatic. The above conditions are believed to be satisfied in the deep interiors of giant planets and in the bulk of the solar convection zone, but  break down in their outermost parts which feature relatively small sound speeds \citep{Ulrich1970,Bahcall1988,Christensen1996,Guillot2004}. The dynamics of these outer layers thus cannot be accounted for within the anelastic framework, and modelers are forced to exclude them from the simulation domain. The dynamical consequences of neglecting these regions remain unclear. 

In summary, both approaches have advantages and drawbacks. While the fully compressible approach is the method of choice for modeling $O(1)$ Mach number flows in near-surface regions of stellar objects, the anelastic approximation seems to be beneficial in the deep interiors where the Mach numbers are usually very small and where the thermodynamic state is close to the adiabat. Unfortunately, in  many astrophysical applications, it remains unclear which approach performs best, with anelastic and fully compressible models being used side by side. 
The main goal of this study is thus twofold: First, we aim to quantify and compare the errors inherent in modeling turbulent convection in both approaches. Secondly, we seek to compare their computational efficiency, thereby guiding modelers in minimizing the tradeoff between accuracy and efficiency for any given situation.

Perhaps somewhat surprisingly, comparing results from the anelastic models currently used in astro- and geophysics to standard fully compressible simulations is non-trivial. This is because the anelastic models usually parameterize the turbulent, subgrid-scale entropy flux, while similar turbulence models are typically not used in fully compressible models. The popularity of turbulence modeling in the anelastic framework stems from the fact that it allows further simplifications of the governing equations, which eases the numerical implementation considerably. Typically, molecular heat conduction is neglected and replaced by an artificial eddy diffusion model that represents turbulent mixing of entropy \citep{Gilman1981,Glatzmaier1984,Braginsky1995,Lantz1999}. This turbulent entropy diffusion model, however, is not mandatory for the actual anelastic approximation, and anelastic equations have been formulated that do not rely on parameterizations of the subgrid-scale transport \citep{Gough1969}. These equations have not found widespread use so far. In order to provide direct comparability between the anelastic and the fully compressible approach, in this study we will restrict ourselves to molecular thermal heat conduction in both cases.

While direct comparisons of anelastic and fully compressible gravity wave dynamics in stably stratified set-ups have been performed in several studies (e.g. \citealp{Davies2003,Klein2010,Brown2012}), the unstable thermal convection case considered in this paper has received less attention so far. The work of \citet{Berkoff2010} focussed on linear magnetoconvection and found good agreement between both approaches for the weakly superadiabatic case. Subsequently, \citet{Lecoanet2014} studied differences between temperature and entropy diffusion, while \citet{Calkins2014b,Calkins2014} focussed on the influence of rotation on the onset of anelastic and fully compressible convection. Their linear study identified shortcomings of the anelastic equations for rapidly rotating, low Prandtl number fluids, where fast density oscillations were found to become non-negligible. 
\citet{Calkins2014b} conclude that fully non-linear studies tracing the validity range of the anelastic approximation are crucial in both rotating and non-rotating systems, especially in the turbulent regime characterized by a broadband frequency spectrum. 
\new{ A first step in this direction has been taken by \citet{Meakin2007}, who compared non-linear anelastic and fully compressible simulations of stellar oxygen burning. The differing physical processes included in each model, however, precluded a one-to-one comparability of the  anelastic and fully compressible influences.}

In this paper, we present the first\new{ systematic} one-to-one comparisons between fully compressible and anelastic numerical simulations of convection in the fully nonlinear, turbulent regime.\new{ As a starting point, we neglect important ingredients of stellar convection, such as spherical geometry, rotation, compositional inhomogeneities, nuclear reactions,  magnetic fields, penetration and overshooting in stably stratified layers, and the corresponding wave-emission. This allows us to quantify the respective errors, as well as the computational efficiency encountered in both approaches in the simplest setup possible. The influences of the above physical processes will be investigated in future studies.}\old{ For simplicity, we focus entirely on non-rotating, non-magnetic convection here. The effects of both rotation and magnetic fields will be discussed in detail in future publications.}

The paper is organized as follows: In section \ref{model}, we start with defining our idealized model, which is followed by discussing the fully compressible equations along with the anelastic approximation in section \ref{comp_an_equations}. A brief overview of the applied numerical methods is given in section \ref{numerics}, while a direct comparison of anelastic and fully compressible results and the computational efficiencies of both approaches are discussed in section \ref{results}. Finally, general conclusions are drawn in section \ref{conclusions}.

\section{MODEL}
\label{model}

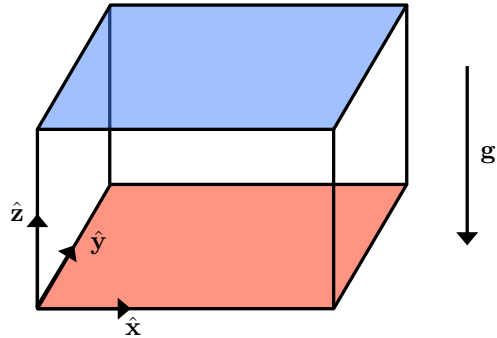
\begin{figure}[tbp]
\setlength{\unitlength}{\linewidth}
\begin{picture}(1.0,0.6)
\put(0.1,0.02){\includegraphics[width=0.8\linewidth]{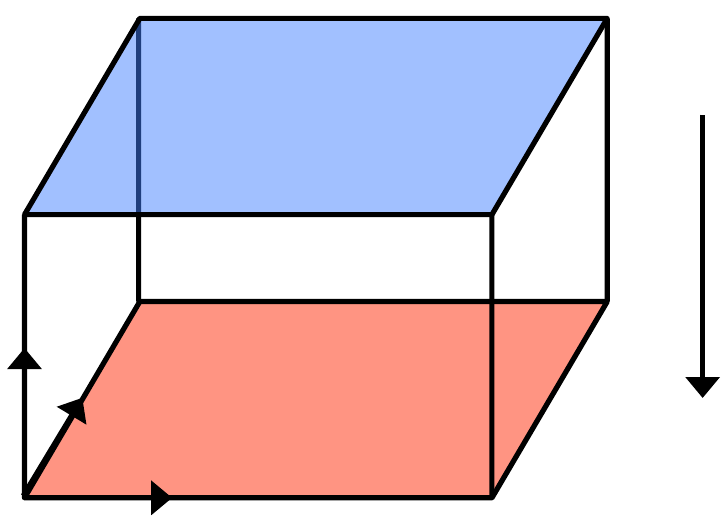}}
\put(0.28,0.00){$\hat{\vec x}$}
\put(0.22,0.15){$\hat{\vec y}$}
\put(0.08,0.20){$\hat{\vec z}$}
\put(0.9,0.31){$\vec g$}
\end{picture}
\caption{Compressible convection is modeled in Rayleigh-B\'enard geometry, i.e. in a Cartesian box that is cooled from above and heated from below. Gravity $\vec g$ is pointing downward, antiparallel to the $z$-axis.}
\label{box_fig}
\end{figure}

Fully compressible and anelastic convection in an ideal gas are modeled in a plane fluid layer of depth $d$ confined between rigid, horizontal plates, as displayed in figure \ref{box_fig}. Gravity $\vec g=-g\hat{\vec z}$ is constant and pointing downward, antiparallel to the unit vector $\hat{\vec z}$. The fluid is cooled from above and heated from below by maintaining a constant, prescribed temperature difference across the layer. The remaining boundary conditions are periodic in the horizontal directions and no slip at the bottom and the top boundary. The ideal gas is characterized by constant dynamic viscosity $\mu=\nu\rho$, heat conductivity $k=c_p\rho\kappa$ and specific heat capacities at fixed volume and pressure, $c_v$ and $c_p$. The quantities $\nu$ and $\kappa$ are the kinematic viscosity and the thermal diffusivity, respectively, which consistently vary across the fluid layer inversely proportional to the density.

The governing equations for fully compressible convection describing the temporal evolution of density $\rho$, temperature $T$, pressure $p$ and velocity $\vec v$ are
\begin{align}
\label{continuity1}
& \partial_{t} \rho + \nabla\cdot(\rho\vec v)=0, \\
\nonumber \\
\nonumber
& \rho\left[\partial_{t} \vec v + (\vec v \cdot \nabla)\vec v\right] = - \nabla p - \rho g \hat{\vec z} \\
\label{momentum1}
& + \mu\left[\nabla^{2}\vec v + \frac{1}{3}\nabla(\nabla\cdot\vec v)\right], \\
\nonumber \\
\nonumber
& c_{v}\rho\left[\partial_{t} T + (\vec v \cdot \nabla) T\right] + p(\nabla\cdot\vec v) = \\
\label{energy1}
& k\nabla^{2}T + 2 \mu \left[e_{ij} - \frac{1}{3}(\nabla\cdot\vec v)\delta_{ij}\right]^{2}, \\
\nonumber \\
\label{state1}
& p=(c_{p}-c_{v})\rho T,
\end{align}

\noindent with $t$ denoting time and $e_{ij}=\frac{1}{2}\left(\partial_{j} v_{i} + \partial_{i} v_{j}\right)$ being the strain rate tensor. Equations (\ref{continuity1}-\ref{energy1}) express the conservation of mass, momentum and energy, respectively, while equation (\ref{state1}) is the ideal gas law.

\section{FULLY COMPRESSIBLE AND AN\-ELAS\-TIC EQUATIONS}
\label{comp_an_equations}

In the following, the anelastic and fully compressible equations are discussed in detail.

\subsection{Reformulation and Non-dimen\-sio\-nali\-zation}
\label{reformulation}

We begin with reformulating the left-hand-side of equation (\ref{energy1}) in the more "anelastic-friendly" way
\begin{align}
\nonumber
& c_{v}\rho\left[\partial_{t} T + (\vec v \cdot \nabla) T\right] + p(\nabla\cdot\vec v) \\
= & c_{p}\rho\left[\partial_{t} T + (\vec v \cdot \nabla) T\right] - \left[\partial_{t} p + (\vec v \cdot \nabla) p\right]
\label{lhs1}
\end{align}
by using equations (\ref{continuity1}) and (\ref{state1}) (details are given in appendix \ref{appendix_volume_work}).
As usual, the thermodynamic quantities are decomposed into a time-independent, vertically varying, hydrostatic and adiabatic background state (index $A$)\footnote{We use the term "adiabatic" here for constant entropy states. More accurately, the background state may be called "isentropic", which however appears to be less common in the literature.} and a superadiabatic part (index $S$), which is allowed to vary in time and space,
\begin{align}
\label{T_A}
& T(t,\vec x) = T_A(z) + T_{S}(t,\vec x), \\
\label{rho_A}
& \rho(t,\vec x) = \rho_A(z) + \rho_{S}(t,\vec x), \\
\label{p_A}
& p(t,\vec x) = p_A(z) + p_{S}(t,\vec x).
\end{align}

\noindent While for subadiabatic or stably stratified fluids a conductive background state is the better choice, the assumption of approximate adiabaticity (i.e. isentropy) is justified in superadiabatic regions, where convection turbulently mixes the fluid and thus homogenizes entropy. 
The background profile can be derived from hydrostaticity $\nabla p=-\rho g \hat{\vec z}$ (i.e. equation (\ref{momentum1}) with $\vec v = 0$) and the thermodynamic relation
\begin{align}
\label{thermodynamics_diff}
\rho T ds = c_{p} \rho dT - \delta_{p}dp,
\end{align}

\noindent with $s$ denoting specific entropy and the dimensionless thermal expansion coefficient being defined as $\delta_{p}=-(\partial \ln \rho / \partial \ln T)$. Note that for an ideal gas, $\delta_p=1$, see (\ref{state1}). By further assuming adiabaticity (i.e. $ds=0$), the background state is found to be 
\begin{align}
& T_A(z) = T_{r}\left(1-\frac{g}{c_{p}T_{r}}z\right), \\
& \rho_A(z) = \rho_{r}\left(1-\frac{g}{c_{p}T_{r}}z\right)^{c_v/(c_p-c_v)}, \\
& p_A(z) = (c_{p} - c_{v}) \rho_{r} T_{r}\left(1-\frac{g}{c_{p}T_{r}}z\right)^{c_p/(c_p-c_v)},
\end{align}

\noindent where the index $r$ in $T_r$ and $\rho_r$ refers to reference values, here defined as the adiabatic values at the bottom boundary. Applying the decomposition of the thermodynamic variables (\ref{T_A}-\ref{p_A}) to equations (\ref{continuity1}-\ref{state1}) results in
\begin{align}
\label{continuity1aa}
& \partial_{t} (\rho_A+\rho_S) + \nabla\cdot\left[(\rho_A+\rho_S)\vec v\right]=0, \\
\nonumber \\
\nonumber
& (\rho_A+\rho_S)\left[\partial_{t} \vec v + (\vec v \cdot \nabla)\vec v\right] = - \nabla (p_A+p_S) \\
\label{momentum1aa}
& - (\rho_A+\rho_S) g \hat{\vec z} + \mu\left[\nabla^{2}\vec v + \frac{1}{3}\nabla(\nabla\cdot\vec v)\right], \\
\nonumber \\
\nonumber
& c_{p}(\rho_A+\rho_S)\left[\partial_{t} (T_A+T_S) + (\vec v \cdot \nabla) (T_A+T_S)\right] \\
\nonumber
& - \left[\partial_{t} (p_A+p_S) + (\vec v \cdot \nabla) (p_A+p_S)\right] = \\
\label{energy1aa}
& k\nabla^{2}(T_A+T_S) + 2 \mu \left[e_{ij} - \frac{1}{3}(\nabla\cdot\vec v)\delta_{ij}\right]^{2}, \\
\nonumber \\
\label{state1aa}
& (p_A+p_S)=(c_{p}-c_{v})(\rho_A+\rho_S) (T_A+T_S).
\end{align}

Within the anelastic approximation, insignificant terms in the above equations are neglected. To judge which terms are negligible, the magnitude of each single term needs to be estimated, which is best done after a proper rescaling. If not stated otherwise, from now on non-dimensional variables will be used. All spatial variables are scaled with the domain depths $d$ and velocity is non-dimensionalized with a convective free-fall velocity $v_f=\sqrt{\Delta\rho g d / \rho_{r}}$. Correspondingly, time is scaled with the free-fall time $t_f=d/v_f=\sqrt{\rho_r d / (\Delta\rho g)}$. In choosing these units, we implicitly assume that shorter timescales, as for example caused by sound waves, play a minor role. 
The scale for temperature $T$ and adiabatic temperature $T_A$ is $T_{r}$, i.e. the temperature at the bottom of the domain, while the superadiabatic temperature difference $\Delta T$, as dictated by the thermal boundary conditions, scales the superadiabatic temperature $T_S$. Since temperature and density perturbations are usually assumed to be closely correlated (see e.g. \citealp{Clayton1968}), the superadiabatic density $\rho_S$ is scaled with the approximate superadiabatic density jump $\Delta \rho=\rho_{r} \Delta T / T_{r}$. Consistently, density $\rho$ and adiabatic background density $\rho_A$ are scaled with $\rho_{r}$, which is the adiabatic density at the bottom of the fluid layer. While pressure $p$ and adiabatic pressure $p_A$ are non-dimensionalized with $(c_{p}-c_{v})\rho_{r}T_{r}$ as suggested by the ideal gas law, the appropriate superadiabatic pressure scale $\Delta\rho g d$ can be inferred from the fact that the superadiabatic pressure $p_S$ extracts kinetic energy from the vertical flows to drive the horizontal motions (see e.g. \citealp{Gough1969}). The non-dimensional thermodynamic quantities\footnote{Note that as the temperature at the bottom of the domain, which is dictated by the boundary conditions, is used to scale the temperature, it follows that $T(z=0)=1$. Therefore, $T_S$ is generally negative for a superadiabatically stratified system as considered here.} then read
\begin{align}
\label{T_A1}
& T(t,\vec x) = T_A(z) + \epsilon T_{S}(t,\vec x), \\
\label{rho_A1}
& \rho(t,\vec x) = \rho_A(z) + \epsilon \rho_{S}(t,\vec x), \\
\label{p_A1}
& p(t,\vec x) = p_A(z) + \epsilon p_{S}(t,\vec x),
\end{align}

\noindent with $\epsilon=\Delta T/T_r$ and the adiabatic background state being
\begin{align}
\label{T_A2}
& T_A(z) = \left(1-Dz\right), \\
\label{rho_A2}
& \rho_A(z) = \left(1-Dz\right)^{1/(\gamma-1)}, \\
\label{p_A2}
& p_A(z) = \left(1-Dz\right)^{\gamma/(\gamma-1)}.
\end{align}


\noindent Upon dividing  equations (\ref{continuity1aa}-\ref{state1aa}) by $\rho_r v_f/d$, $\rho_r g$, $c_p\rho_r v_f T_r/d$, and $(c_p-c_v)\rho_r T_r$, respectively, we obtain
\begin{align}
\label{continuity1a}
& \epsilon\partial_{t} \rho_{S} + \nabla\cdot\left[(\rho_A + \epsilon\rho_{S})\vec v\right]=0, \\
\nonumber \\
\nonumber
& \epsilon(\rho_A+\epsilon\rho_{S})\left[\partial_{t} \vec v + (\vec v \cdot \nabla)\vec v\right] = \\
\nonumber
& - \nabla \left(\frac{1-\frac{1}{\gamma}}{D}p_A + \epsilon p_{S}\right) - (\rho_A + \epsilon\rho_{S}) \hat{\vec z} \\
\label{momentum1a}
& + \epsilon\sqrt{\frac{Pr}{Ra}}\left[\nabla^{2}\vec v + \frac{1}{3}\nabla(\nabla\cdot\vec v)\right], \\
\nonumber \\
\nonumber
& (\rho_A+\epsilon\rho_{S})\left[\epsilon\partial_{t} T_{S} + (\vec v \cdot \nabla) (T_A + \epsilon T_{S})\right] \\
\nonumber
& - \left\{\epsilon D \partial_{t} p_{s} + (\vec v \cdot \nabla) \left[\left(1-\frac{1}{\gamma}\right) p_A + \epsilon D p_{s}\right]\right\} \\
\nonumber
& = \frac{1}{\sqrt{Ra Pr}} \nabla^{2}(T_A + \epsilon T_S) \\
\label{energy1a}
& + 2\epsilon D\sqrt{\frac{Pr}{Ra}} \left[e_{ij} - \frac{1}{3}(\nabla\cdot\vec v)\delta_{ij}\right]^{2}, \\
\nonumber \\
\label{state1a}
&p_A +\epsilon\frac{D}{1-\frac{1}{\gamma}}p_S=(\rho_A+\epsilon\rho_S)(T_A+\epsilon T_S).
\end{align}

\noindent Due to the non-dimensionalization with characteristic scales, all variables and operators are $O(1)$ and the magnitude of each term in equations (\ref{T_A1}-\ref{state1a}) can be estimated by its prefactor. The non-dimensional control parameters  $\epsilon$, $Ra$, $Pr$, $\gamma$, and $D$ occurring in these coefficients are discussed in the following section.

\subsection{Control Parameters and Magnitude of the Terms}
\label{parameters}

Seven control parameters determine the fate of the convection system governed by (\ref{T_A2}-\ref{p_A2}) and (\ref{continuity1a}-\ref{state1a}). The superadiabaticity
\begin{equation}
\epsilon=\frac{\Delta T}{T_{r}}=\frac{\Delta \rho}{\rho_{r}}
\label{eps}
\end{equation}

\noindent compares the superadiabatic temperature difference as dictated by the boundary conditions to a typical reference temperature that is---as all other reference values---evaluated at the bottom. We will show later that $\epsilon$ constrains the typical Mach number $M$, which is defined as the ratio of a typical convective free-fall velocity and the speed of sound. The Rayleigh number
\begin{equation}
Ra=\frac{gd^{3}\Delta T}{\nu_{r}\kappa_{r} T_{r}}=\frac{gd^{3}\epsilon}{\nu_{r}\kappa_{r}}
\label{Ra}
\end{equation}

\noindent controls the vigor of convection with large $g$, $d$, and $\epsilon$ enhancing and large diffusivities $\nu$ and $\kappa$ reducing the convective vigor. More formally $Ra$ is the ratio of the product of the diffusive timescales $d^2/\kappa_r ~ d^2/\nu_r$ to the square of the free-fall timescale $t_f^2$. The Prandtl number
\begin{equation}
Pr=\frac{\nu}{\kappa},
\label{Pr}
\end{equation}

\noindent which for the setup chosen here is constant with depth, denotes the ratio of momentum diffusivity to the thermal diffusivity and therefore is a material property. It effectively controls the importance of inertia in the system, with $Pr\ll 1$ leading to strong and $Pr\gg 1$ leading to weak inertial effects. The ratio of the heat capacities defines the parameter
\begin{equation}
\gamma=\frac{c_{p}}{c_{v}},
\label{gamma}
\end{equation}

\noindent while the Dissipation number
\begin{equation}
D=\frac{gd}{c_{p}T_{r}}
\label{Diss}
\end{equation}

\noindent can be interpreted in several different ways. Its name originates from the fact that it constraints how much internal energy can be generated by viscous dissipation, i.e. $D$ is a measure for the significance of viscous heating with $0\le D \le 1$. This becomes evident from (\ref{eps}) and (\ref{gamma}), which allow to rearrange the dissipation number to $D=\frac{1}{\gamma}\frac{gd\Delta\rho}{\rho_{r}c_{v}\Delta T}=\frac{1}{\gamma}\frac{E_{pot}}{\Delta E_{int}}$. This alternative formulation reveals that the dissipation number is proportional to the ratio of the available potential energy $E_{pot}=gd\Delta\rho$, which drives convection, to the typical internal energy variations $\Delta E_{int}=\rho_{r}c_{v}\Delta T$. As viscous heating results from the dissipation of convective kinetic energy (for which $E_{pot}$ defines the upper limit), viscous heating can only significantly contribute to internal energy variations if $E_{pot}$ is of the same order of magnitude as $\Delta E_{int}$. 
$D$ can also be interpreted to be the inverse adiabatic temperature scale height evaluated at the bottom boundary. 
Finally, the dissipation number is directly linked to the density contrast $\chi$ that may serve as an alternative parameter. It is defined as the ratio of the adiabatic density at the bottom and at the top,
\begin{equation}
\chi=\frac{\rho_A^{bot}}{\rho_A^{top}}=\frac{\rho_A(z=0)}{\rho_A(z=1)}=(1-D)^{-1/(\gamma-1)}.
\label{chi_D}
\end{equation}

The total mass of the fluid, as determined by the initial conditions, and the aspect ratio of the periodic box form the last two control parameters.

The scaled equations (\ref{continuity1a}-\ref{state1a}), which still represent the full compressible dynamics, can be further simplified by noting that the $\epsilon^0$ terms in equations (\ref{momentum1a}-\ref{state1a}) balance exactly. In the momentum equation (\ref{momentum1a}), the $\epsilon^0$ terms simply represent the hydrostatic balance of the reference state, i.e. $-(1-1/\gamma)/D\nabla p_A - \rho_A\hat{\vec z}=0$. Likewise, the first two $\epsilon^0$ terms in the energy equation (\ref{energy1a}) $\rho_A v_z \partial_z T_A - (1-1/\gamma) v_z \partial_z p_A=0$ balance because of (\ref{T_A2}-\ref{p_A2}). Note that the conduction term $1/\sqrt{RaPr}\nabla^2 T_A$ drops out here because the adiabatic temperature gradient is constant in our simple model configuration\footnote{For general depth dependent heat conductivities $k$ and adiabatic temperature gradients $\nabla T_A$, this term must be retained. It then effectively acts as a heat source or sink and drives or hinders convection with the magnitude being estimated by the term's prefactor $1/\sqrt{RaPr}$. For astrophysical systems that exhibit large Rayleigh numbers this magnitude is typically very small and may be comparable or even smaller than the magnitude of the $\epsilon^1$ terms representing the usual convective perturbation. For numerical simulations that do not reach realistic parameter values, the diffusion of adiabatic background temperature, however, may be of significance.}. Finally, in equation (\ref{state1a}), the $\epsilon^0$ terms $p_A=\rho_A T_A$ just represent the ideal gas law for the reference state. 

By subtracting the $\epsilon^0$ terms from  (\ref{momentum1a}-\ref{state1a}) and dividing by $\epsilon$, we arrive at
\begin{align}
\label{continuity2}
& \epsilon\partial_{t} \rho_{S} + \nabla\cdot\left[(\rho_A + \epsilon\rho_{S})\vec v\right]=0, \\
\nonumber \\
\nonumber
& (\rho_A+\epsilon\rho_{S})\left[\partial_{t} \vec v + (\vec v \cdot \nabla)\vec v\right] = - \nabla p_{S} \\
\label{momentum2}
& - \rho_{S} \hat{\vec z} + \sqrt{\frac{Pr}{Ra}}\left[\nabla^{2}\vec v + \frac{1}{3}\nabla(\nabla\cdot\vec v)\right], \\
\nonumber \\
\nonumber
& (\rho_A+\epsilon\rho_{S})\left[\partial_{t} T_{S} + (\vec v \cdot \nabla) T_{S}\right] -D \rho_{s} v_{z} \\
\nonumber
& - D \left[\partial_{t} p_{s} + (\vec v \cdot \nabla) p_{s}\right] = \frac{1}{\sqrt{Ra Pr}} \nabla^{2}T_S \\
\label{energy2}
& + 2D\sqrt{\frac{Pr}{Ra}} \left[e_{ij} - \frac{1}{3}(\nabla\cdot\vec v)\delta_{ij}\right]^{2}, \\
\nonumber \\
\label{state2}
&\frac{D}{1-\frac{1}{\gamma}}\frac{p_{S}}{p_A} = \frac{T_{S}}{T_A} + \frac{\rho_{S}}{\rho_A} + \epsilon \frac{\rho_{S}}{\rho_A} \frac{T_{S}}{T_A}
\end{align}

\noindent which describe fully compressible convection as perturbations from the adiabatic, hydrostatic background state.

\subsection{Anelastic Approximation}

The energy conserving anelastic equations, that can also be derived more formally by a rigorous amplitude expansion (e.g. \citealp{Gough1969,Lantz1999}), follow from (\ref{continuity2}-\ref{state2}) in the limit $\epsilon\rightarrow 0$, resulting in

\begin{align}
\label{continuity3}
& \nabla\cdot\left(\rho_A\vec v\right)=0, \\
\nonumber \\
\nonumber
& \rho_A\left[\partial_{t} \vec v + (\vec v \cdot \nabla)\vec v\right] = - \nabla p_{S} \\
\label{momentum3}
& - \rho_{S} \hat{\vec z} + \sqrt{\frac{Pr}{Ra}}\left[\nabla^{2}\vec v + \frac{1}{3}\nabla(\nabla\cdot\vec v)\right], \\
\nonumber \\
\nonumber
& \rho_A\left[\partial_{t} T_{S} + (\vec v \cdot \nabla) T_{S}\right] -D \rho_{s} v_{z} \\
\nonumber
& - D \left[\partial_{t} p_{s} + (\vec v \cdot \nabla) p_{s}\right] = \frac{1}{\sqrt{Ra Pr}} \nabla^{2}T_S \\
\label{energy3}
& + 2D\sqrt{\frac{Pr}{Ra}} \left[e_{ij} - \frac{1}{3}(\nabla\cdot\vec v)\delta_{ij}\right]^{2}, \\
\nonumber \\
\label{state3}
&\frac{D}{1-\frac{1}{\gamma}}\frac{p_{S}}{p_A} = \frac{T_{S}}{T_A} + \frac{\rho_{S}}{\rho_A}.
\end{align}
Note that for the setup chosen here, the superadiabaticity parameter $\epsilon$ drops out of the non-dimensional anelastic equations\footnote{For the general case that contains the diffusion of background temperature, the $\epsilon$-parameter controls the significance of this process and is thus retained.}. Furthermore, the well-known Boussinesq equations describing shallow convection follow in the limit  $D\rightarrow0$.

In a very simple manner the above equations illustrate the neglected physical processes in the anelastic and the Boussinesq approximation: The continuity equation (\ref{continuity2}) reveals that by letting $\epsilon\rightarrow 0$, sound waves are effectively filtered out as the time derivative term becomes negligible. Furthermore, unpleasant nonlinearities disappear in  (\ref{momentum2}-\ref{state2}). In the Boussinesq limit $D\rightarrow 0$, the energy equation (\ref{energy2}) uncovers that pressure loses its role in the energy budget, while viscous heating can be neglected as the available potential energy is much smaller than internal energy variations (c.f. section \ref{parameters}). Equation (\ref{state2}) further shows that the superadiabatic density is directly proportional to the superadiabatic temperature in the Boussinesq limit. Finally, the Mach number
\begin{equation}
M=\frac{v_f}{v_s}=\frac{\sqrt{\Delta\rho g d / \rho_r}}{\sqrt{c_p(c_p-c_v)T_r/c_v}}=\sqrt{\frac{\epsilon D}{\gamma - 1}},
\label{machnumber}
\end{equation}

\noindent based on the free-fall velocity $v_f$ and the speed of sound $v_s$ at the bottom of the domain, can be estimated from the input parameters. Obviously, it is considered to be small in both, the anelastic and the Boussinesq approximation. Note that the Mach number can serve as an alternative control parameter that replaces $\epsilon$. In solar and giant planets' interiors, where $D$ and $\gamma - 1$ can typically be assumed to be $O(1)$, the square of the Mach number is crudely approximated by the superadiabaticity,
\begin{align}
M^2\approx \epsilon,
\end{align}
which suggests that the anelastic approximation holds for $M\ll 1$.

\section{NUMERICAL REALIZATION}
\label{numerics}

The equations governing fully compressible convection (\ref{continuity2}-\ref{state2}) are solved on a collocated grid using second order finite differences and a third order upwind method for the advection terms. A semi-implicit time stepping scheme based on a third order Adams-Bashforth / backward-difference formula (AB3/BDF3) is applied (e.g. \citealp{Boyd2001,Peyret2002}). All terms except for the vertical diffusion terms are treated explicitly.

The anelastic simulations that will be presented in this paper are performed with an anelastic code, which is a modified version of the Boussinesq code by \citet{Stellmach2008}. It uses a mixed pseudo-spectral  fourth order finite-difference discretization of the spatial derivatives and an AB3/BDF3 time integration scheme, which treats all linear terms implicitly. Instead of using (\ref{continuity2}-\ref{state2}) directly, for numerical reasons it turns out to be beneficial to use an equivalent formulation based on entropy rather than temperature. The relevant equations  (\ref{anelasticcontinuity}-\ref{anelasticenergy}) are derived in detail in appendix \ref{anelastic}.

\section{RESULTS}
\label{results}

\begin{figure*}[tbp]
\setlength{\unitlength}{\linewidth}
\begin{picture}(1.0,0.32)
\put(0.0,0.0){\includegraphics[width=0.48\linewidth]{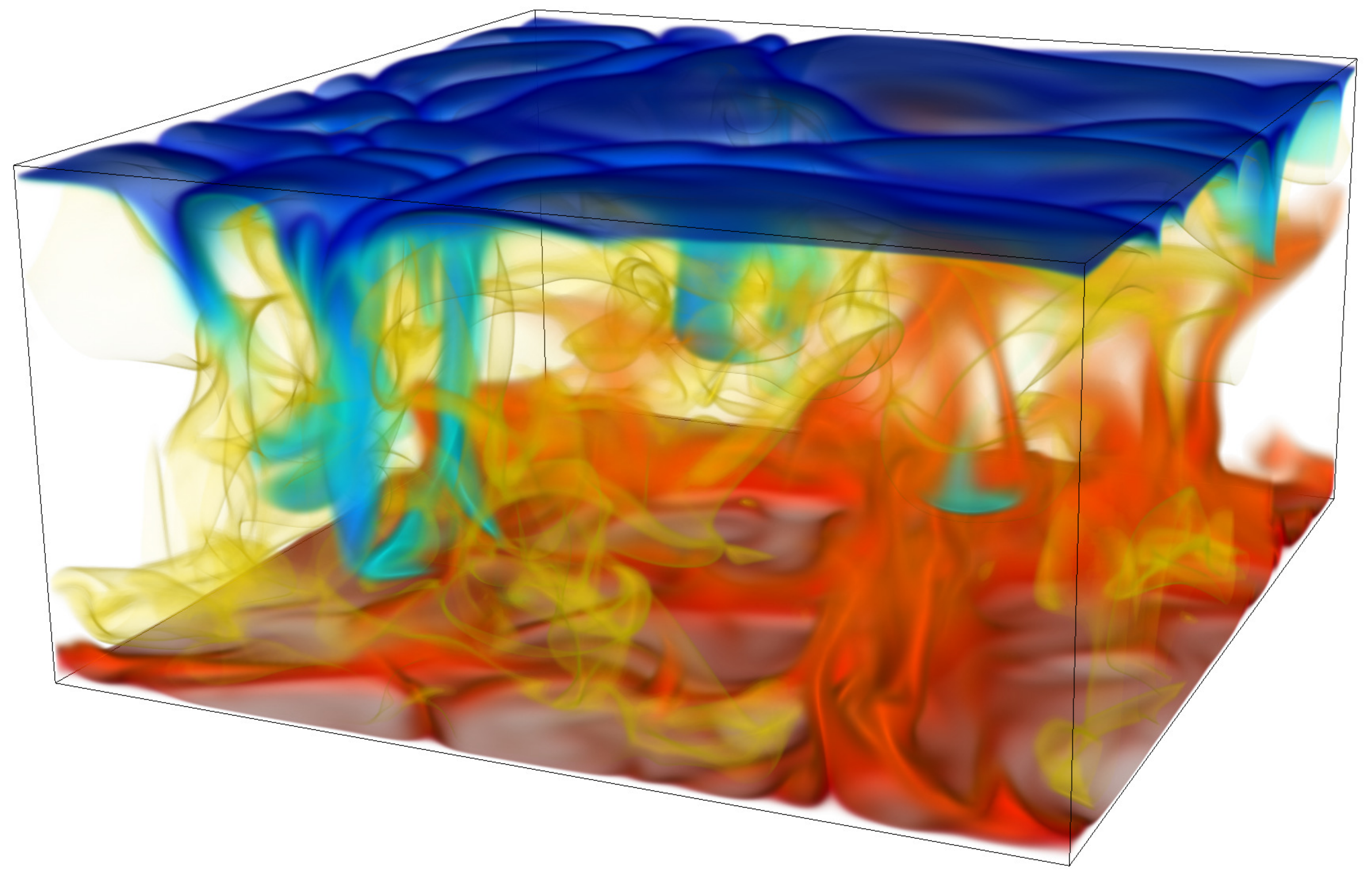}}
\put(0.52,0.0){\includegraphics[width=0.48\linewidth]{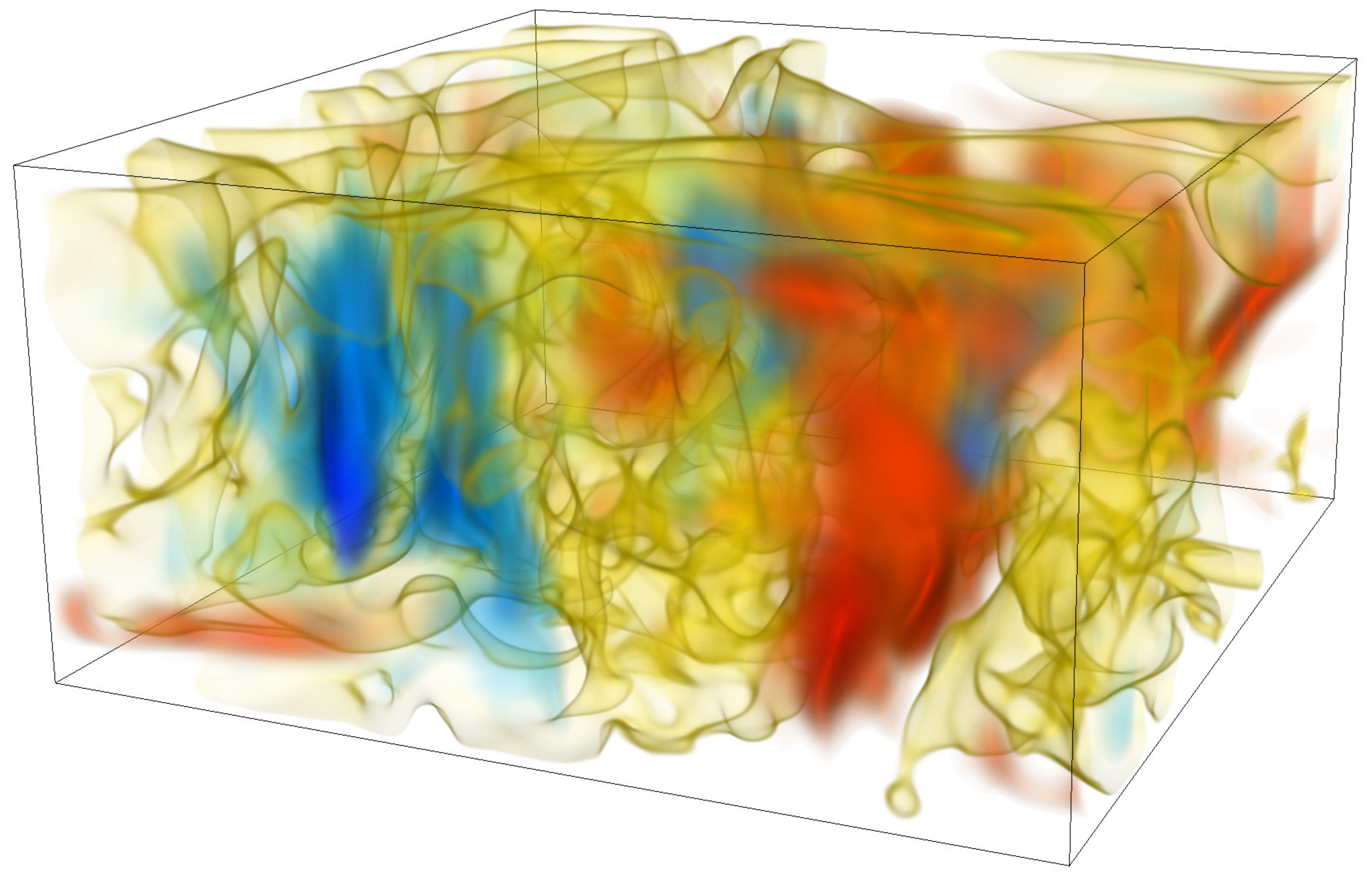}}
\put(0.0,0.28){a)}
\put(0.52,0.28){b)}
\end{picture}
\caption{Typical volume renderings of the superadiabatic temperature $T_S$ (a) and vertical velocity $v_z$ (b) for an anelastic simulation run that reached statistical equilibrium. Red colors denote warm, buoyant material and positive $v_z$, blue signifies cold fluid and negative $v_z$, and yellow structures refer to intermediate values of $T_S$ and $v_z$. The corresponding parameters are $\epsilon=0$, $Ra=10^7$, $Pr=0.7$, $\gamma=5/3$ and $\chi=\exp(1)\approx 2.72$. Corresponding snapshots taken from numerical simulations of fully compressible convection look qualitatively the same and cannot be visually distinguished from the displayed example. A stereoscopic 3-d version of these volume renderings, which reflects the full 3-d structures when wearing red-cyan filter glasses, is shown in figure \ref{anaglyph} in the appendix.}
\label{snapshot_fig}
\end{figure*}

In this section results from a suite of anelastic and fully compressible\new{ direct} numerical simulations\new{ (DNS)} are presented in order to test the accuracy and efficiency of both approaches in the fully nonlinear regime of convection.

\begin{table}
\centering
\begin{tabular}{>{$}c<{$} >{$}c<{$} >{$}c<{$} >{$}c<{$} >{$}c<{$} >{$}c<{$}}
\hline
\epsilon & Ra & \chi & \text{Resolution} & t_{run} & Re \\
\hline
0 & 10^4 & 2.72 & 144^2 \times 129 & 519 & 25.0 \\
0.01 & 10^4 & 2.72 & 128^3  &146 & 25.0 \\
0.05 & 10^4 & 2.72 & 128^3  & 326 & 25.6 \\
0.1 & 10^4 & 2.72 &  128^3  & 463 & 26.3 \\
0.15 & 10^4 & 2.72 &  128^3  & 148 & 27.0 \\
0.2 & 10^4 & 2.72 &  128^3  & 151 & 27.7 \\
0.25 & 10^4 & 2.72 &  128^3  & 77.8 & 28.4 \\
0.3 & 10^4 & 2.72 &  128^3  & 185 & 29.1 \\
0.35 & 10^4 & 2.72 &  128^3  & 92.0 & 30.0 \\
0.4 & 10^4 & 2.72 &  128^3  & 248 & 30.9 \\
\hline
0 & 10^5 & 2.72 &  144^2 \times 129 & 4180 & 99.7 \\
0.1 & 10^5 & 2.72 &  128^3  & 3519 & 102 \\
0.2 & 10^5 & 2.72 &  128^3  & 2990 & 104 \\
0.3 & 10^5 & 2.72 &  128^3  & 2937 & 107 \\
0.4 & 10^5 & 2.72 &  128^3  & 4260 & 111 \\
\hline
0 & 10^6 & 2.72 &  192^2 \times 193 & 3003 & 316 \\
0.1 & 10^6 & 2.72 &  192^3  & 1364 & 322 \\
0.2 & 10^6 & 2.72 &  192^3  & 1063 & 330 \\
0.3 & 10^6 & 2.72 &  192^3  & 2384 & 339 \\
0.4 & 10^6 & 2.72 &  192^3  & 2041 & 350 \\
\hline
0 & 10^7 & 2.72 &  288^2 \times 257 & 1913 & 954 \\
0.1 & 10^7 & 2.72 &  256^3 & 1373 & 973 \\
0.2 & 10^7 & 2.72 &  256^3  & 1262 & 991 \\
0.3 & 10^7 & 2.72 &  256^3  & 1878 & 1016 \\
0.4 & 10^7 & 2.72 &  256^3  & 1066 & 1050 \\
\hline
0 & 10^6 & 4.48 &  192^2 \times 193 & 2744 & 300 \\
0.1 & 10^6 & 4.48 &  192^3 & 1205 & 313 \\
0 & 10^6 & 7.39 &  192^2 \times 193 & 2535 & 294 \\
0.1 & 10^6 & 7.39 &  192^3  & 1422 & 299 \\
0 & 10^6 & 12.18 &  192^2 \times 193 & 2811 & 280 \\
0.1 & 10^6 & 12.18 &  192^3  & 1347 & 286 \\
0 & 10^6 & 20.1 &  192^2 \times 193 & 2945 & 269 \\
0.1 & 10^6 & 20.1 &  192^3  & 1464 & 271 \\
\hline
\end{tabular}
\caption{\new{Overview of the simulations carried out for this study, with $Pr=0.7$ and $\gamma=5/3$ applying to all simulations. The horizontal dimensions of the simulation domain are $l_x=l_y=2d$, resulting in an aspect ratio of two. While the spatial resolution is given in the number of x, y, and z grid points, $t_{run}$ denotes the run time measured in free-fall times, and $Re=v_{rms}/\sqrt{Pr/Ra}$ is the approximated Reynolds number, with the non-dimensional root-mean-square velocity $v_{rms}$ being defined in equation (\ref{vrms_def}). While all $Ra=10^4$ cases result in stationary solutions, the remaining simulations stay time-dependent.}}
\label{simulation_table}
\end{table}

Equations (\ref{continuity2}-\ref{state2}) are solved for various superadiabaticities ($0\le \epsilon \le 0.4$), density contrasts ($2.72\approx \exp(1) \le \chi \le \exp(3) \approx 20.1$ or analogously $0.49 \le D \le 0.86$) and Rayleigh numbers ($10^4\le Ra \le 10^7$). \new{See Table \ref{simulation_table} for an overview of all simulations.} The simulation runs with $\epsilon=0$ are carried out with the anelastic code, while the remaining simulations are executed with our independent code for fully compressible convection as described in section \ref{numerics}. The remaining four control parameters are kept constant for all simulations, with the Prandtl number set to $Pr=0.7$ and the ratio of specific heats chosen to represent a monoatomic ideal gas, $\gamma=\frac{5}{3}$. The horizontal dimensions of the simulation domain are $l_x=l_y=2d$, resulting in an aspect ratio of two. The total mass of the fluid is determined by the initial state, for which we choose the hydrostatic, conductive solution with
\begin{align}
\nonumber
T(t=0,z)= & T_A + \epsilon T_S(t=0) \\
= & \left[1- (D+\epsilon)z\right].
\end{align}

\noindent The integral over the corresponding initial density distribution
\begin{align}
\nonumber
\rho(t=0,z)= & \rho_A + \epsilon\rho_{S}(t=0) \\
= & \left[1-(D+\epsilon)z\right]^n,
\label{dens}
\end{align}
where
\begin{equation}
n=\frac{\gamma}{\gamma-1}\frac{D}{D+\epsilon}-1
\label{polytropic_index}
\end{equation}
is the polytropic index, determines the total mass. Note that the polytropic index $n$ is often used as an alternative parameter to the superadiabaticity $\epsilon$ (e.g. \citealp{Brummell1996,Brummell1998,Berkoff2010}). For $\chi\approx2.72$ and $0.1\le \epsilon \le 0.4$, which are typical parameters for this study, the polytropic index varies within the range $1.07\ge n \ge 0.37$.

To give the reader a feeling for the level of turbulence reached in our simulations, figure \ref{snapshot_fig} shows a typical snapshot of an anelastic simulation run that reached statistical equilibrium. Corresponding snapshots taken from numerical simulations of fully compressible convection look qualitatively similar.

\subsection{Comparison of fully compressible and anelastic results}

\begin{figure*}[tbp]
\includegraphics[width=\linewidth]{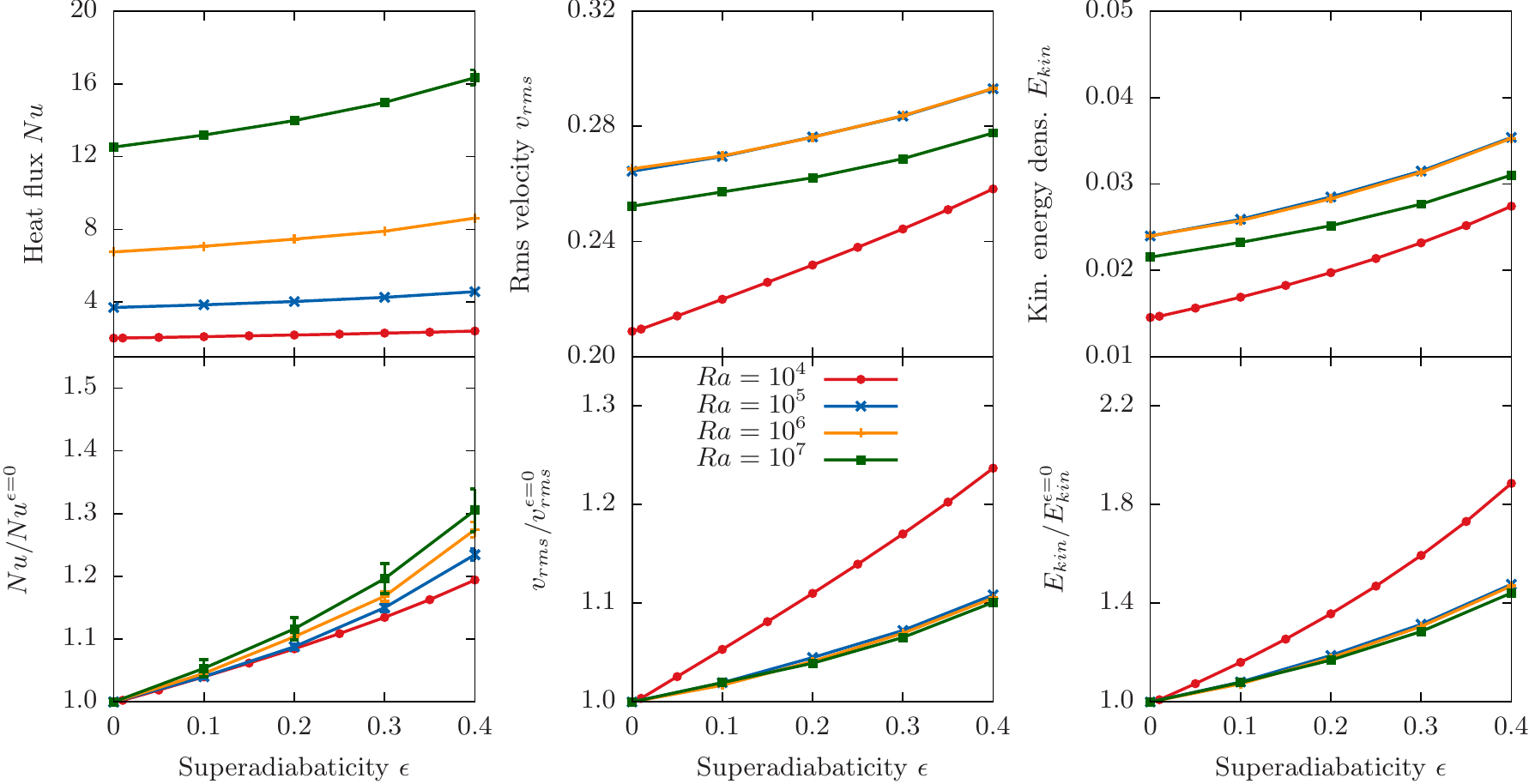}
\caption{Global diagnostic quantities are plotted against the superadiabaticity parameter $\epsilon$. From left to right, the graphs in the top row display the heat flux in terms of a Nusselt number $Nu$, the root mean square velocity $v_{rms}$, and the kinetic energy density $E_{kin}$. The bottom row shows the same quantities normalized to the corresponding anelastic values ($\epsilon=0$). For all Rayleigh numbers, the fully compressible results converge to the anelastic values for $\epsilon \rightarrow 0$. For large Rayleigh numbers $Ra$ and superadiabaticities $\epsilon$ smaller than $0.3$, the outputs from compressible convection deviate by no more than $30\%$ from the associated anelastic values. The error bars given for the Nusselt numbers are estimates based on the difference between temporally averaged Nusselt numbers computed at the top and bottom boundary. In all cases, $\chi=2.72$.}
\label{diagnostics_fig}
\end{figure*}

Global diagnostic quantities can provide a first impression as to what extent the anelastic approximation holds. Initially, we vary $\epsilon$ and $Ra$, while keeping $\chi=\exp(1)\approx 2.72$ constant. Covering several orders of magnitude in Rayleigh number $Ra$, figure \ref{diagnostics_fig} shows three different global diagnostics plotted against the superadiabaticity parameter $\epsilon$. From left to right, the graphs in the top row display the heat flux in terms of the Nusselt number
\begin{equation}
Nu=\left. - \partial_z \bar T_S \right|_{z=0},
\end{equation}
the root-mean-square velocity
\begin{equation}
v_{rms} = \left<\sqrt{\left<\vec v^2\right>_{v}}\right>_{t},
\label{vrms_def}
\end{equation}
and the kinetic energy density
\begin{equation}
E_{kin} = \frac{1}{2}\left<\rho\vec v^2\right>_{t,v},
\end{equation}
where brackets $\left<...\right>$ denote temporal (index $t$), volume (index $v$) and/or horizontal (index $h$) averages, while an overbar implies both temporal and horizontal averaging, $\bar{...}=\left<...\right>_{t,h}$.
The bottom row shows the same quantities normalized by the corresponding anelastic values $Nu^{\epsilon=0}$, $v_{rms}^{\epsilon=0}$ and $E_{kin}^{\epsilon=0}$.

As expected, the fully compressible cases converge to the anelastic results as $\epsilon$ is decreased. From  theoretical considerations (cf. section \ref{comp_an_equations}), we expect the convergence to be linear in $\epsilon$, a trend which is most clearly seen for $Ra=10^{4}$, where the flow field is stationary. For the fluctuating solutions encountered at larger $Ra$, higher order terms appear to contribute considerably to the dynamics even for moderate superadiabaticities $\epsilon \gtrsim 0.3$, where the linear scaling is observed to break down.

Another important question is how the ratio of fully compressible and anelastic results scales with Rayleigh number for a fixed value of $\epsilon$. Interestingly, while both $v_{rms}/v_{rms}^{\epsilon=0}$ and $E_{kin}/E_{kin}^{\epsilon=0}$ decrease with $Ra$, the relative heat flux $Nu/Nu^{\epsilon=0}$ increases, without showing any sign of convergence over the range of Rayleigh numbers studied. Whether $Nu/Nu^{\epsilon=0}$ converges to a finite value beyond $Ra=10^7$ or continues to increase monotonically is left to future investigations. The answer is of great importance for astrophysical systems, which typically have Rayleigh numbers much larger than those considered here. 

In summary, perhaps the most important conclusion to be drawn from figure \ref{diagnostics_fig}  is that in the turbulent, high Rayleigh number regime, the difference between the fully compressible results and the corresponding anelastic values remain moderate in all cases studied. For $\epsilon \lesssim 0.3$, the relative deviations are less than $10\%$, $20\%$, and $30\%$ for $\epsilon=0.1$, $\epsilon=0.2$, and $\epsilon=0.3$, respectively. As a crude rule of thumb, $\epsilon$ thus provides a reasonable estimate of the relative error.


\begin{figure*}[tbp]
\includegraphics[width=\linewidth]{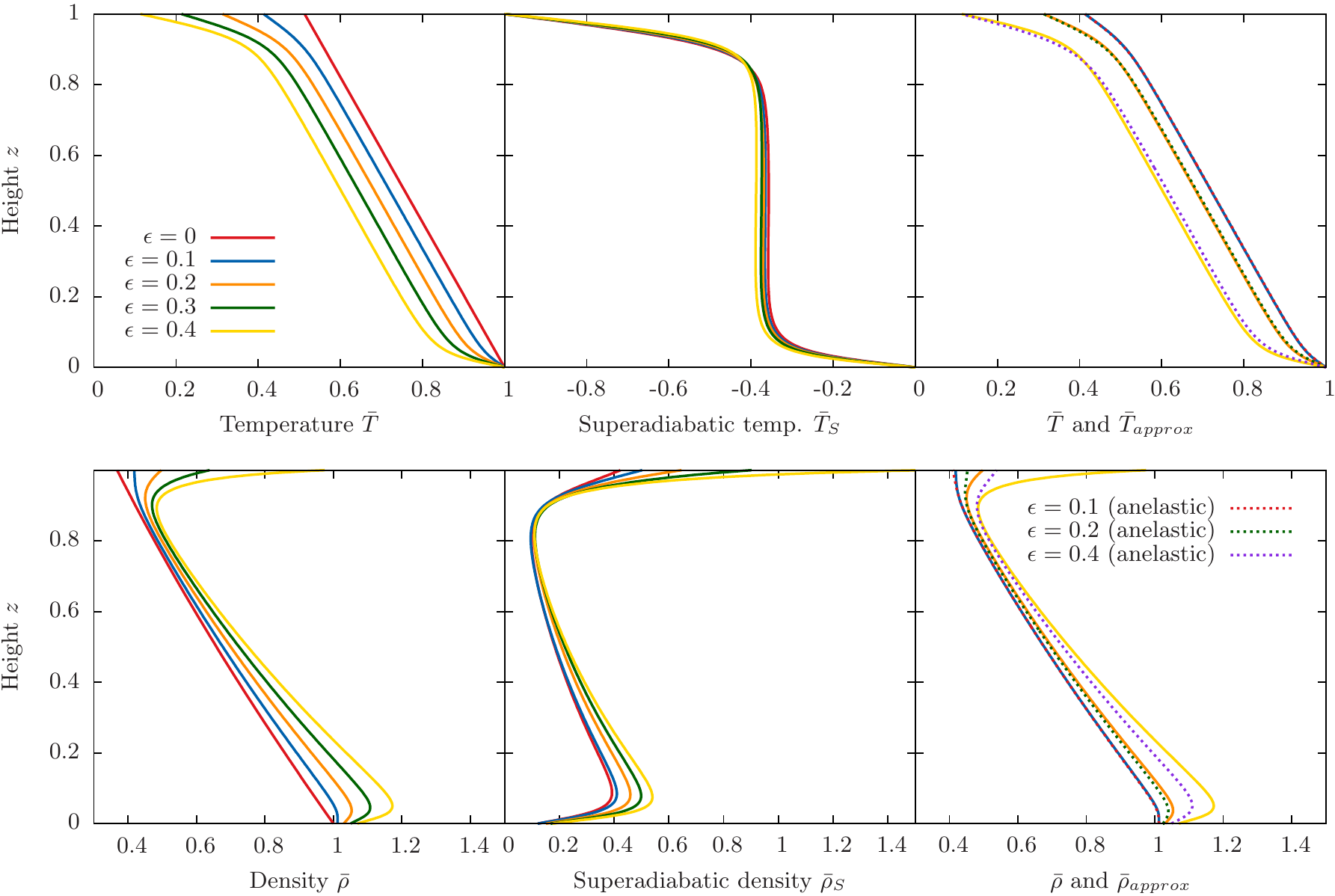}
\caption{Depth profiles of temperatures and densities for various superadiabaticities $\epsilon$ for the $Ra=10^6$ case. The top row, from left to right, displays plots of total temperatures $\bar T=T_A+\epsilon \bar T_S$, superadiabatic temperatures $\bar T_S$, and comparisons of $\bar T$ with approximated total temperatures $\bar T_{approx}=T_A+\epsilon \bar T_S^{\epsilon=0}$. These are reconstructed from the anelastic limit case by extrapolating to finite superadiabaticities $\epsilon$. Analogously, in the bottom row the total densities $\bar \rho=\rho_A+\epsilon\bar \rho_S$, superadiabatic densities $\bar \rho_S$, and the approximated densities $\bar \rho_{approx}=\rho_A+\epsilon\bar \rho_S^{\epsilon=0}$ in comparison with $\bar \rho$ are presented in the panels from left to right. The profiles for $\bar T$, $\bar T_S$, $\bar \rho$, and $\bar \rho_S$ obtained from compressible convection simulations converge to those of the anelastic case as $\epsilon$ decreases. No deviations between the approximated temperature profiles $\bar T_{approx}$ and the corresponding unapproximated ones $\bar T$ can be seen by eye for $\epsilon\le 0.2$, while for $\epsilon=0.4$ the deviations lie within a few percent. Deficiencies of the anelastic approximation become evident when comparing $\bar \rho$ and $\bar \rho_{approx}$. While the $\epsilon\le 0.2$ cases still fit very well, larger superadiabaticities seem to be problematic especially near the top boundary.}
\label{dens_temp_fig}
\end{figure*}

After discussing global diagnostic quantities and their variations with $\epsilon$, a more detailed view is provided by vertical profiles obtained from horizontal and temporal averages of the solutions. Figure \ref{dens_temp_fig} shows temperature and density profiles for $Ra=10^6$ and various values of $\epsilon$. The top row, from left to right, displays plots of total temperature $\bar T=T_A+\epsilon \bar T_S$, superadiabatic temperature $\bar T_S$ and comparisons of $\bar T$ with an approximated total temperature $\bar T_{approx}=T_A+\epsilon \bar T_S^{\epsilon=0}$. Analogously, the total density $\bar \rho=\rho_A+\epsilon\bar \rho_S$, superadiabatic density $\bar \rho_S$ and a comparison of the approximated density $\bar \rho_{approx}=\rho_A+\epsilon \bar \rho_S^{\epsilon=0}$ with $\bar \rho$ are presented in the bottom panels. Note that the profiles of $\bar T$ and $\bar \rho$ for $\epsilon=0$ in the left column simply represent the adiabatic background state. 

Just like the global diagnostic quantities, the profiles for $\bar T$, $\bar T_S$, $\bar \rho$, and $\bar \rho_S$ obtained from compressible convection simulations converge to those of the anelastic case as $\epsilon$ decreases. When comparing the approximated temperature profiles $\bar T_{approx}$ with the corresponding $\bar T$, no difference can be seen by eye for $\epsilon\le 0.2$, while for $\epsilon=0.4$ the deviations lie within a few percent. Possible deficiencies of the anelastic approximation become evident when comparing $\bar \rho$ and $\bar \rho_{approx}$. While the $\epsilon\le 0.2$ cases still fit very well, larger superadiabaticities seem to be problematic especially near the top boundary, where the deviations almost reach $100\%$ for the $\epsilon=0.4$ case.

\begin{figure*}[tbp]
\includegraphics[width=\linewidth]{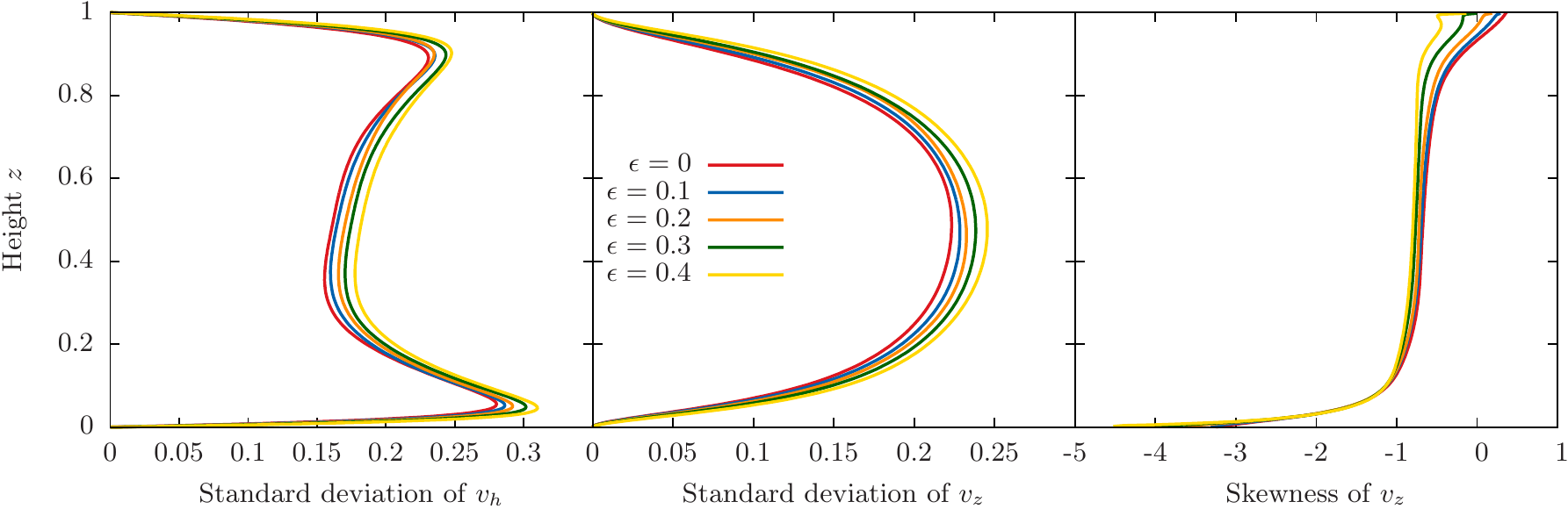}
\caption{The panels from left to right show the standard deviation of the horizontal velocity $\sigma_{v_h}$, the standard deviation of the vertical velocity $\sigma_{v_z}$, and the skewness of the vertical velocity $\gamma_{v_z}$. All profiles obtained from compressible convection simulations converge against those of the anelastic case ($\epsilon=0$) as $\epsilon$ decreases.}
\label{velo_fig}
\end{figure*}

Further depth profiles are shown for the velocity field in figure \ref{velo_fig}. The panels from left to right show the time averaged standard deviation of the horizontal velocity $\sigma_{v_h}$, the time averaged standard deviation of the vertical velocity $\sigma_{v_z}$ and the time averaged skewness of the vertical velocity $\gamma_{v_z}$, where the definitions 
\begin{align}
\sigma_X(z)= & \left<\sqrt{\left[X - \left< X \right>_h \right]^2}\right>_t, \\
\gamma_X(z)= & \left< \frac{\left[X-\left< X \right>_h \right]^3}{\left[X - \left< X \right>_h \right]^2}\right>_t
\end{align}
are used for the time averaged standard derivation and skewness of a quantity $X$. In agreement with the results presented above, all profiles obtained from compressible convection simulations converge to those of the anelastic case as $\epsilon$ decreases.

\begin{figure*}[tbp]
\includegraphics[width=\linewidth]{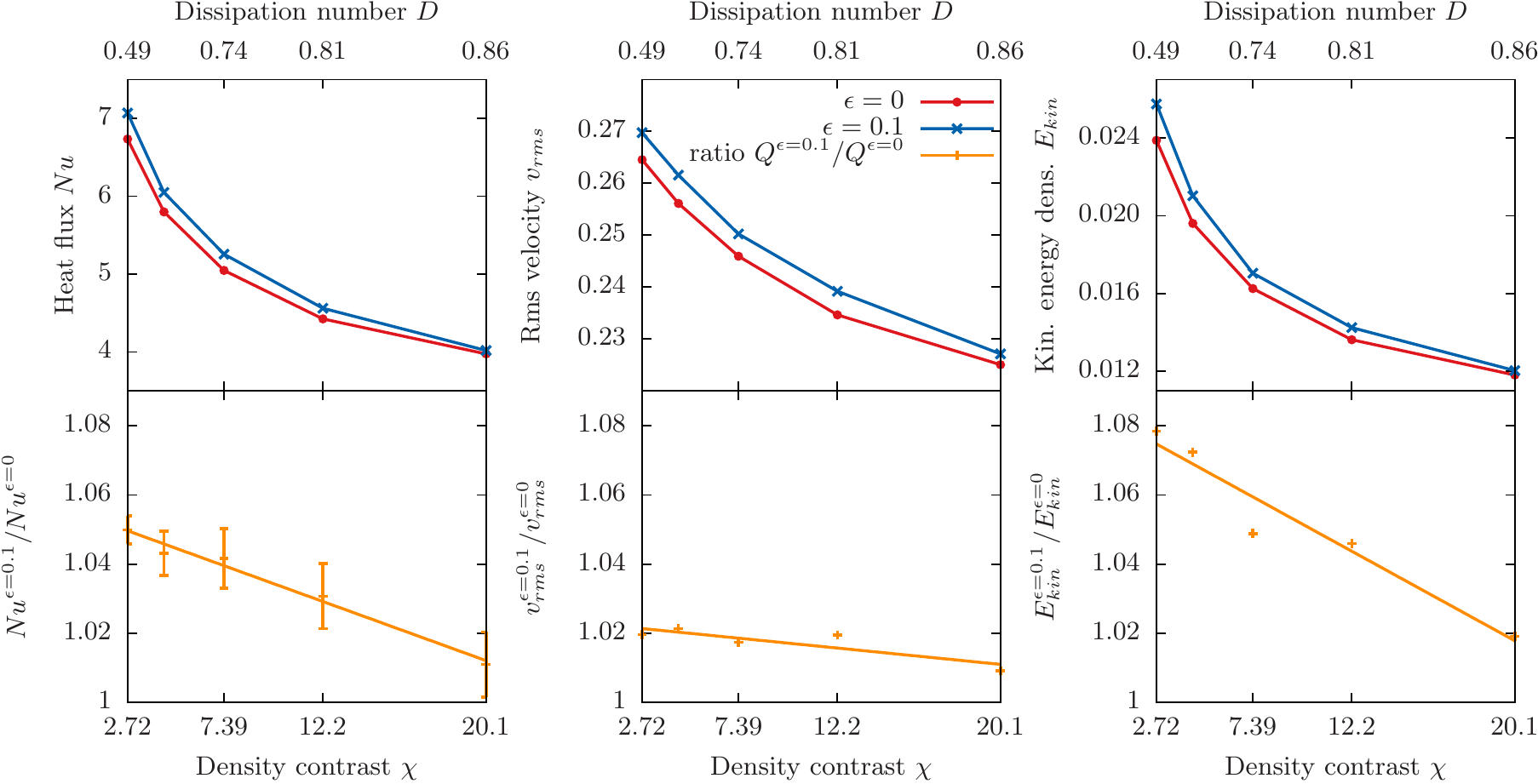}
\caption{Global diagnostic quantities obtained from fully compressible ($\epsilon=0.1$) and anelastic ($\epsilon=0$) simulations ($Ra=10^6$) are plotted against the density contrast $\chi$ (and Dissipation number $D$) in the top row. From left to right, the graphs display the heat flux in terms of a Nusselt number $Nu$, root mean square velocity $v_{rms}$, and kinetic energy density $E_{kin}$. The bottom row shows the ratio of the respective quantities $Q^{\epsilon=0.1}/Q^{\epsilon=0}$. To guide the eye, a linear fit is also plotted revealing that the relative differences of compressible and anelastic outputs generally decrease with increasing density contrasts. The error bars given for the ratio of the Nusselt number result from differences in the time-averaged bottom and top Nusselt numbers.}
\label{dens_contrast1_fig}
\end{figure*}

\begin{figure*}[tbp]
\includegraphics[width=\linewidth]{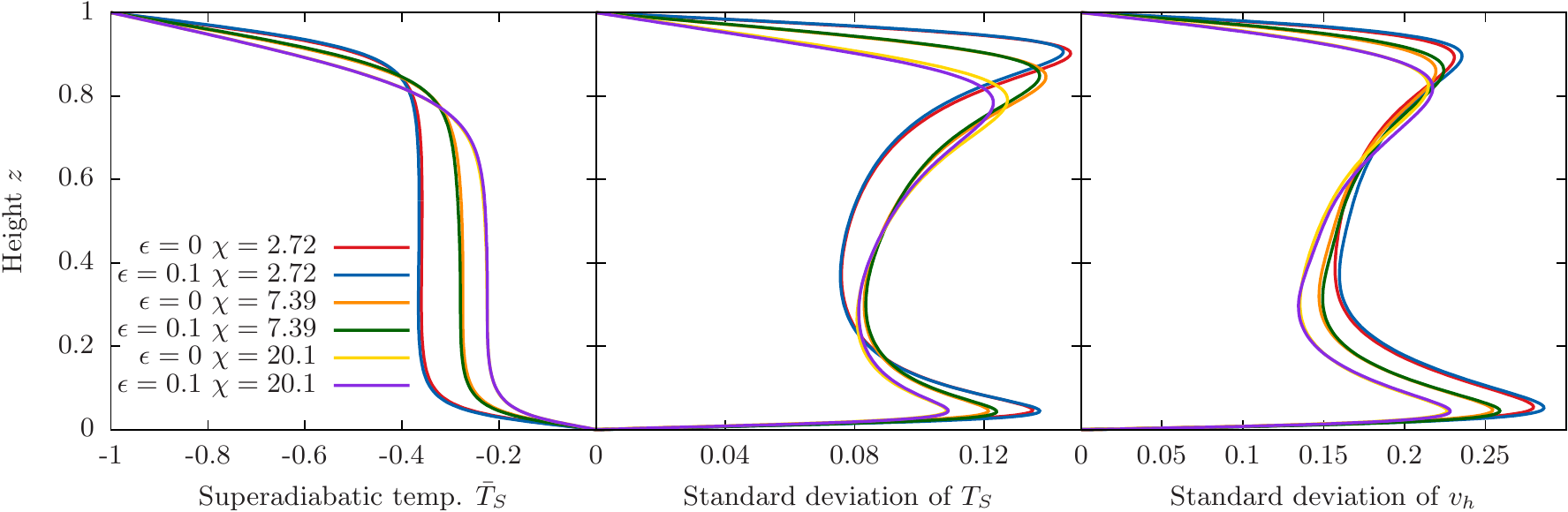}
\caption{Depth profiles of compressible ($\epsilon=0.1$) and anelastic ($\epsilon=0$) convection are shown for different density contrasts $\chi$ with constant $Ra=10^6$. The panels from left to right show the mean superadiabatic temperature $\bar T_S$, the standard deviation of the superadiabatic temperature $\sigma_{T_S}$, and the standard deviation of the horizontal velocity $\sigma_{v_h}$. In the lower half of the fluid container, which covers the largest amount of the fluid's mass, the deviations between the fully compressible and the anelastic case decrease with increasing density contrast, while no distinct trend can be seen in the top part.}
\label{dens_contrast2_fig}
\end{figure*}

Finally, we focus on the influence of the density contrast $\chi$ on the accuracy of the anelastic approximation by comparing anelastic ($\epsilon=0$) and fully compressible ($\epsilon=0.1$) simulations for $Ra=10^6$ and $2.72 \le \chi \le 20.1$. In analogy to figure \ref{diagnostics_fig}, figure \ref{dens_contrast1_fig} displays the ratio of compressible and anelastic diagnostic outputs. Shown are the Nusselt number, root-mean-square velocity and kinetic energy density for various density contrasts. Interestingly, the differences between anelastic and fully compressible diagnostics decrease with increasing density contrast. 

The plots shown in figure \ref{dens_contrast2_fig} provide further insight into this nonlinear effect. The panel on the left shows depth profiles of the superadiabatic temperature. They reveal that the magnitude of $\bar T_S$ in the lower part of the fluid container, which contains the bulk of the fluid's mass, decreases as the density contrast increases. This implies that increasing the density contrast $\chi$ effectively decreases the superadiabatic perturbations in the bulk region. As the accuracy of the anelastic approximation is proportional to the relative magnitude of the superadiabatic perturbations, increasing the density contrast effectively increases the precision of the anelastic equations. 

This view is further supported by the two panels on the right of figure \ref{dens_contrast2_fig} that display the depth profiles of the standard deviation of the superadiabatic temperature $\sigma_{T_S}$ and the standard deviation of the horizontal velocity $\sigma_{v_h}$. Both exhibit pronounced maxima near the boundaries that mark the edges of the thermal and viscous boundary layers. The case with the biggest density contrast $\chi=20.1$ clearly reveals that the differences between anelastic and compressible profiles are largest near the location of the top maxima, in accordance with the relatively large magnitude of the superadiabatic temperature within the upper thermal boundary layer. Near the bottom boundary, where the superadiabatic perturbations are small, the anelastic curve cannot be visually distinguished from the compressible profile.

\subsection{Computational Efficiency}

After discussing the quantitative differences between fully compressible and anelastic results, we briefly turn to the question of which approach is computationally more efficient.

The main computational benefit of employing the anelastic approximation is that sound waves are filtered out and thus do not need to be resolved (e.g. \citealp{Gough1969,Glatzmaier1984,Lantz1999}). Assuming that the timestep length in simulations of anelastic convection is constrained by the free-fall velocity $v_f$, while in the fully compressible case the limit is set by the sound speed $v_s$, we expect
\begin{align}
\frac{\Delta t_{max}^{compress}}{\Delta t_{max}^{anelastic}} \approx& \frac{v_f}{v_s} = M \nonumber \\
 =& \sqrt{\frac{\epsilon D}{\gamma - 1}} = \sqrt{\frac{\epsilon (1-\chi^{1-\gamma})}{\gamma - 1}},
 \label{eq:time_step_ratio}
\end{align}
where (\ref{machnumber}) and (\ref{chi_D}) have been used. The Mach number $M$, or, alternatively, for fixed $\chi$ and $\gamma$, the superadiabaticity $\epsilon$ is therefore expected to control the time step ratio.

\begin{figure}[tbp]
\includegraphics[width=\linewidth]{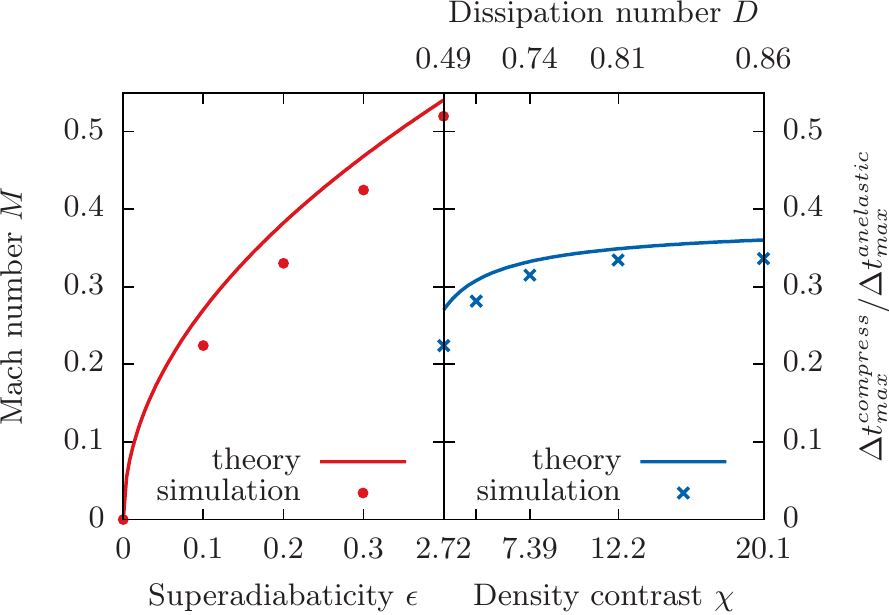}
\caption{
Plot of the theoretically and numerically found Mach numbers $M$ for a monoatomic ideal gas, i.e. $\gamma=5/3$. Numerical $M$ are obtained from simulations at $Ra=10^6$ and $Pr=0.7$ and show good agreement with theoretical predictions that are generally slightly larger. While the left panel (red) shows a plot of $M$ against superadiabaticity $\epsilon$ for a constant density contrast of $\chi=2.72$, the right panel (blue) displays $M$ against $\chi$ (or alternatively $D$) for a fixed $\epsilon=0.1$. The Mach number can be interpreted to be the ratio of the respective maximum timestep lengths of compressible and anelastic convection $\Delta t_{max}^{compress} / \Delta t_{max}^{anelastic}$, which allows to compare the theoretical efficiency of anelastic and compressible numerical codes.}
\label{machnumber_fig}
\end{figure}

In order to check the validity of the above estimate, figure \ref{machnumber_fig} compares the Mach numbers 
\begin{equation}
M^{\text{num}} := \frac{\max( |{\bf v}| )}{v_s(z=0)}
\end{equation}
obtained from numerical simulations with the theoretical prediction (\ref{eq:time_step_ratio}). While the maximum is taken over the entire computational domain, the sound speed is evaluated at the bottom boundary, where it becomes largest.  Both $\epsilon$ and $\chi$  are varied for a mono-atomic ideal gas with $\gamma=5/3$. As expected, the free-fall estimate (\ref{eq:time_step_ratio})  slightly overestimates the real Mach numbers, but overall the behavior is captured reasonably well.

In general, $\epsilon$ is the most important parameter controlling the relative efficiency of both approaches. As figure \ref{machnumber_fig} shows, for $\epsilon=0.1$, Mach numbers around $0.3$ are reached, such that anelastic codes can use time steps which are roughly three times larger than those possible in fully compressible simulations. This however is not guaranteed to result in real computational savings, as individual time steps tend to be more costly in anelastic simulations. In contrast to the fully compressible case, the pressure field adapts instantaneously and is governed by an elliptic equation, which complicates the time stepping procedure. Although efficient solution techniques are readily available from the extensive literature on incompressible computational fluid dynamics, it is not unreasonable to assume that the additional costs slow down the computation of a single time step by a factor of three, such that both approaches reach similar efficiency. Indeed, we experienced that some of our compressible simulations at $\epsilon=0.1$ were computationally more efficient than their anelastic counterparts.

The above conclusions have been drawn solely from simulations of turbulent, compressible Ray\-leigh-B\'enard convection. Effects occurring from strong rotation or magnetic fields may alter the picture significantly. For example, if thin boundary layers need to be resolved, the time step restriction arising from the sound waves may become prohibitive in fully compressible simulations. This situation could arise for example in planetary cores, where the presence of rigid walls combined with the rapid rotation generates very thin Ekman boundary layers, which nevertheless appear to be dynamically active \citep{Stellmach2014} and thus need to be accounted for.

We finish this section by noting that more detailed efficiency comparisons are beyond the scope of this paper. Much will depend on the numerical algorithms employed, on the degree to which the codes are tuned to the machine they run on, and on the architecture of the computer itself.

\section{CONCLUSIONS}
\label{conclusions}

This paper has presented the first one-to-one comparison of anelastic and fully compressible turbulent convection. Our goal was to quantify the accuracy and efficiency of both methods for the simple test case of turbulent Rayleigh-B\'enard convection in an ideal gas. 


The relation between the anelastic and fully compressible equations has been carved out in detail, without invoking subgrid-scale turbulence modeling at any stage. The anelastic approximation is expected to hold in the limit of small superadiabaticity $\epsilon$, such that the Mach number  $M \sim \sqrt{\epsilon}$  remains small. We have shown that the fully compressible equations can be manipulated into a particular non-dimensional form, which consists of terms representing the anelastic dynamics plus $O(\epsilon)$ correction terms that guarantee the fully compressible physics. In the limit $\epsilon \rightarrow 0$ these correction terms vanish and the usual anelastic equations, as rigorously derived by formal amplitude expansions  in previous works (e.g. \citealp{Gough1969,Lantz1999}), are recovered. Our approach helps to make the relation between the anelastic and fully compressible equations fully transparent, and also reveals that the familiar Boussinesq equations result in the double limit $\epsilon \rightarrow 0, D \rightarrow 0$, where $D=gd/(c_p T_r)$ is the dissipation number. The requirement of small Dissipation number is equivalent to a shallow convective system with a depth that is much smaller than the typical temperature scale height.

A key aspect of this work was to quantify the differences between fully compressible and anelastic results. Therefore a suite of anelastic and fully compressible numerical simulations of thermal convection has been carried out. We compared global diagnostic quantities as well as depth profiles of the most important statistical moments of thermodynamic variables and velocities. All simulations reveal a coherent picture, showing that the fully compressible results converge to the anelastic ones with decreasing $\epsilon$. The relative deviation between both cases was found to be approximately equal to the superadiabaticity $\epsilon$, indicating linear convergence as predicted by theory. For $\epsilon \gtrsim 0.3$ this linear trend is broken and larger deviations are encountered. Besides depending on the superadiabaticity, the degree to which both approaches give consistent results is controlled by the density contrast of the system, i.e. the ratio of bottom to the top density. Interestingly, due to a nonlinear effect, larger density contrasts reduce the quantitative differences between anelastic and fully compressible models in our simulations.

A further aspect of this work dealt with the comparison of the numerical efficiency of the anelastic and the fully compressible approach. In cases with $\epsilon \ll 1$ it is usually argued that solving the anelastic equations is computationally more efficient than solving their fully compressible counterparts, because numerically costly sound waves are filtered out. While our results generally confirm this argument, they also show that fully compressible models appear to become more efficient than anelastic simulations for $\epsilon \ge O(0.1)$.

The implications of our results for the simulations of astrophysical flow phenomena might be illustrated by considering the specific example of solar convection. Standard solar models like, for example, Model S \citep{Christensen1996}\footnote{Model S is available online at \url{http://astro.phys.au.dk/~jcd/solar_models/}}, allow to estimate the superadiabaticity $\epsilon(z)=\left[\partial_z T - (\partial_z T)_A\right]/\partial_z T$ as a function of depth within the solar interior. The superadiabaticity is predicted to be many orders of magnitude smaller than one in the lower $99\%$ of the convection zone. It only reaches $O(1)$ values within the outermost one per cent, close to the solar photosphere. This suggests that results obtained using the anelastic equations are indeed highly accurate in models excluding the thin outermost layer where the approximation breaks down. The dynamical consequences of neglecting this layer, however, need further investigation.

In contrast, the fully compressible approach in principle is capable of capturing the relevant physical processes throughout the entire convection zone. This, however, forces modelers to use unrealistically large values for $\epsilon$ in the bulk of the convection zone for numerical reasons. An important result of our study is that this procedure introduces moderate errors only. Even for $\epsilon \approx 0.1$, where fully compressible codes tend to become more efficient than anelastic models, the error in global diagnostics such as the overall heat transport or the average kinetic energy was found to be of the order of $10 \%$. The impact on the turbulent flow statistics was also shown to remain modest. In comparison to other sources of errors, arising for example from the inability to reach a realistic turbulence level in numerical simulations, a ten percent error seems tolerable. 

The above conclusions have been drawn from numerical simulations that neglect important ingredients of stellar convection, such as\new{ spherical geometry, rotation, compositional inhomogeneities, nuclear reactions,  magnetic fields, penetration and overshooting in stably stratified layers, the corresponding wave-emission}, and of course they did not reach the extreme flow conditions of the solar interior.\new{ While boundary layers play an important role in regulating the convection efficiency in the simulations presented here \citep{Grossmann2000,Petschel2013}, their dominance is less evident in a more realistic model involving much higher Rayleigh numbers \citep{Kraichnan1962,Spiegel1971,Grossmann2000} and more realistic boundary conditions \citep{Brummell2002}.} Applying our results to the Sun is therefore somewhat speculative\new{ and the inclusion of additional physical processes in future comparative studies is clearly desirable}.  In particular, an issue that might arise in rapidly rotating systems has recently been pointed out by \citet{Calkins2014b}, who argued that the anelastic approximation breaks down in the geo- and astrophysically relevant case of rapid rotation and low Prandtl number. A study similar to the one presented here, but including the effects of rapid rotation, is needed to resolve this question and is currently underway.



\acknowledgments

\section*{ACKNOWLEDGMENTS}
The computations have been carried out on the PALMA computer cluster at M\"unster University and on the supercomputer JUQUEEN at the Forschungszentrum J\"ulich. This work was supported by the the German Research Foundation under the Priority Program 1488  (Planetary Magnetism).






\appendix

\section*{APPENDIX}

\section{VOLUME WORK TERM AND ENERGY EQUATION}

\label{appendix_volume_work}

The volume work term $p(\nabla\cdot\vec v)$ in equation (\ref{energy1}) is based on the divergence of the velocity field, which is problematic in the anelastic approach. In the anelastic approximation the velocity field is constrained by the anelastic continuity equation $\nabla\cdot(\rho_A \vec v)=0$, which misses the information of the small superadiabatic density changes driving convection. \citet{Spiegel1960} deal with a related problem when deriving the Boussinesq approximation for shallow convection in an ideal gas. They show that $p(\nabla\cdot\vec v)$ is non-negligible, as small superadiabatic density variations become important, although the Boussinesq continuity equation requires incompressibility, i.e. $\nabla\cdot\vec v=0$.

Following their procedure, the volume work term in the fully compressible energy equation (\ref{energy1}) can be reformulated by using the full continuity equation (\ref{continuity1}) and the ideal gas law (\ref{state1}). In order to do this, both (dimensional) equations are reorganized as follows,
\begin{align}
\partial_{t}\rho+\nabla\cdot(\rho\vec v) =  0 & \Longrightarrow \nabla\cdot\vec v =  - \frac{1}{\rho}\left[\partial_{t}\rho + (\vec v \cdot\nabla)\rho\right], \\
p=(c_{p}-c_{v})\rho T & \Longrightarrow d\rho=\frac{dp}{(c_{p}-c_{v})T} - \frac{\rho}{T}dT.
\end{align}

\noindent Using the above expressions, the volume work term in (\ref{energy1}) can be formulated in terms of temperature and pressure, rather than with the divergence of the velocity field,
\begin{align}
\nonumber
p(\nabla\cdot\vec v) = & - \frac{p}{\rho}\left[\partial_{t}\rho + (\vec v \cdot\nabla)\rho\right] \\
= & (c_{p}-c_{v})\rho\left[\partial_{t} T + (\vec v \cdot \nabla) T\right] - \left[\partial_{t}p + (\vec v \cdot \nabla)p\right].
\label{anelastic_volume_work}
\end{align}

\noindent The anelastic expression for the left hand-side of the energy equation then can be derived by decomposing the thermodynamic variables in an adiabatic and a superadiabatic part, as described in section \ref{reformulation}, and by neglecting all terms involving nonlinearities of variables denoting the superadiabatic part,
\begin{align}
\nonumber
& (c_{p}-c_{v})\rho\left[\partial_{t} T + (\vec v \cdot \nabla) T\right] - \left[\partial_{t}p + (\vec v \cdot \nabla)p\right] \\
\nonumber
= & (c_{p}-c_{v})(\rho_A+\rho_S)\left[\partial_{t} T_S + (\vec v \cdot \nabla) (T_A+T_S)\right] - \left[\partial_{t}p_S + (\vec v \cdot \nabla)(p_A+p_S)\right] \\
\overset{\text{anelastic}}{=} & (c_{p}-c_{v})\rho_A\left[\partial_{t} T_S + (\vec v \cdot \nabla) (T_A+T_S)\right] +(c_p-c_v)\partial_z T_A v_z \rho_S - \left[\partial_{t}p_S + (\vec v \cdot \nabla)(p_A+p_S)\right].
\end{align}

\noindent The anelastic energy equation then results in
\begin{align}
\nonumber
& c_p\rho_A\left[\partial_{t} T_S + (\vec v \cdot \nabla) T_S\right] +c_p\partial_z T_A v_z \rho_S - \left[\partial_{t}p_S + (\vec v \cdot \nabla)p_S\right] \\
= & k\nabla^{2}(T_A+T_S) + 2 \mu \left[e_{ij} - \frac{1}{3}(\nabla\cdot\vec v)\delta_{ij}\right]^{2}.
\label{energy_appendix}
\end{align}

In contrast, \citet{Rogers2005} and \citet{Glatzmaier2014} derive a different version of the volume work term that is not equivalent to ours. Instead of using the full continuity equation and the full ideal gas law and finally making the anelastic approximation (i.e. neglecting all terms that are proportional to nonlinearities of superadiabatic thermodynamic quantities), they apply the anelastic versions of both equations,
\begin{align}
\nabla\cdot(\rho_A\vec v)=0 & \Longrightarrow \nabla\cdot \vec v=-\frac{\partial_z\rho_A}{\rho_A}v_z \\
\frac{p_S}{p_A}=\frac{\rho_S}{\rho_A}+\frac{T_S}{T_A} & \Longrightarrow p_S=(c_p-c_v)(T_A\rho_S+\rho_AT_S).
\end{align}

\noindent which, when utilized for the volume work term, directly reveals their anelastic version
\begin{align}
\nonumber
p(\nabla\cdot\vec v) = & - (p_A+p_S)\frac{\partial_z\rho_A}{\rho_A}v_z \\
= & - (c_p-c_v) \left[T_A\partial_z\rho_A + \frac{T_A}{\rho_A}\partial_z\rho_A \rho_S + \partial_z\rho_A T_S\right]v_z.
\end{align}

\noindent The anelastic energy equation from the \citet{Rogers2005,Glatzmaier2014} point of view, then results in
\begin{align}
c_v\rho_A\left[\partial_{t} T_S + (\vec v \cdot \nabla) T_S\right] -(c_p-c_v)\partial_z \rho_A v_z  T_S = & k\nabla^{2}(T_A+T_S) + 2 \mu \left[e_{ij} - \frac{1}{3}(\nabla\cdot\vec v)\delta_{ij}\right]^{2}.
\label{rogers_energy_appendix}
\end{align}

\noindent This anelastic energy equation differs from ours (\ref{anelastic_volume_work}) and essentially involves the superadiabatic temperature as the only time-dependent thermodynamic variable. This handy formulation, however, has the disadvantage that corresponding numerical simulations carried out by us neither showed conservation of energy nor matched results with fully compressible numerical simulations.

\section{EQUATIONS GOVERNING FULLY COMPRESSIBLE CONVECTION IN ENTROPY FORMULATION}

For some applications the entropy formulation of the energy equation (\ref{energy1}), which, at this point, is given in dimensional form
\begin{align}
\rho T \left[\partial_{t} s + (\vec v \cdot \nabla) s\right] = k\nabla^{2}T + 2 \mu \left[e_{ij} - \frac{1}{3}(\nabla\cdot\vec v)\delta_{ij}\right]^{2}
\label{entropy_equation}
\end{align}

\noindent might be favorable. It can be derived by applying the (dimensional) thermodynamic relation for entropy
\begin{align}
\label{thermodynamics_diff}
\rho T ds = c_{p} \rho dT - \delta_{p}dp.
\end{align}

\noindent When assuming $\delta_p=1$, as valid for an ideal gas, the integration of equation (\ref{thermodynamics_diff}) 
reveals the integrated form of the thermodynamic relation for entropy, which directly relates entropy to temperature and pressure
\begin{align}
\label{thermodynamics1}
& s - s_r = c_p \ln \frac{T}{T_r} - (c_p-c_v) \ln \frac{p}{p_r},
\end{align}

\noindent where $s_r$ is the reference entropy evaluated at the bottom of the domain. Decomposing the thermodynamic variables as done in section \ref{reformulation} and exploiting that the dimensional adiabatic background entropy profile reads
\begin{align}
s_A=c_p \ln\frac{T_A}{T_r}-(c_p-c_v)\ln\frac{p_A}{p_r} + s_r=s_r,
\label{thermo_background}
\end{align}

\noindent allows for the reformulation of the whole set of governing equations (\ref{continuity1}-\ref{state1}) in terms of entropy instead of temperature. The non-dimensional forms of the entropy formulation of the energy equation (\ref{entropy_equation}) and the thermodynamic relation for entropy (\ref{thermodynamics1}) can be derived by using (\ref{thermo_background}) and $\Delta s= c_p \Delta T / T_r$ to scale the superadiabatic entropy,
\begin{align}
(\rho_A+\epsilon\rho_{S})(T_A+\epsilon T_{S})\left[\partial_{t} s_{S} + (\vec v \cdot \nabla) s_{S}\right] = \frac{1}{\sqrt{Ra Pr}} \nabla^{2}T_S + 2D\sqrt{\frac{Pr}{Ra}} \left[e_{ij} - \frac{1}{3}(\nabla\cdot\vec v)\delta_{ij}\right]^{2},
\label{energy2_entropy}
\end{align}
\begin{align}
s_S = \frac{1}{\epsilon} & \left[\ln\left(1+\epsilon\frac{T_S}{T_A}\right) - \left(1-\frac{1}{\gamma}\right)\ln\left(1+\frac{\epsilon D}{1-\frac{1}{\gamma}}\frac{p_S}{p_A}\right)\right].
\label{thermodynamics2}
\end{align}

\section{EQUATIONS GOVERNING ANELASTIC CONVECTION IN ENTROPY FORMULATION}
\label{anelastic}

The anelastic equations in entropy formulation appear from (\ref{continuity2}-\ref{momentum2}), (\ref{state2}) and (\ref{energy2_entropy}-\ref{thermodynamics2}) in the limit $\epsilon=0$. In order to arrive at the entropy formulation of the anelastic equations the thermodynamic relation for entropy (\ref{thermodynamics2}) needs to be derived for the limit case $\epsilon\rightarrow 0$. As $\lim\limits_{x\rightarrow 0}[\ln(1+a x)/x]=a$ the integrated form of the thermodynamic relation for entropy in the limit $\epsilon=0$ results in
\begin{align}
\label{thermodynamics_appendix}
s_S = \frac{T_S}{T_A} - D \frac{p_S}{p_A}.
\end{align}

\noindent When now replacing $T_{S}$ in the anelastic ($\epsilon=0$) ideal gas law (\ref{state2}), the superadiabatic density reads
\begin{align}
\rho_{S}=\frac{D}{\gamma-1} \frac{p_{S}}{T_A} - \rho_A s_{S}.
\label{rhoS}
\end{align}

\noindent In the following we will express all dynamically varying thermodynamic variables in terms of $s_{S}$ and $p_{S}$.

\subsection{Pressure and buoyancy term}

Applying the Lantz-Braginsky-Roberts trick \citep{Lantz1999,Braginsky1995} to the anelastic ($\epsilon=0$) momentum equation (\ref{momentum2}) rearranges the pressure and buoyancy terms
\begin{align}
\nonumber
-\frac{1}{\rho_A}\nabla p_{S} - \frac{\rho_{S}}{\rho_A}\hat{\vec z} = & -\frac{1}{\rho_A}\nabla p_{S} - \left(\frac{D}{\gamma-1}\frac{p_{S}}{\rho_AT_A} - s_{S}\right) \hat{\vec z} \\
\nonumber
= & -\nabla\frac{p_{S}}{\rho_A} - \left( \frac{\partial_{z}\rho_A}{\rho_A^{2}}p_{S} + \frac{D}{\gamma - 1}\frac{p_{S}}{\rho_AT_A} -  s_{S}\right) \hat{\vec z} \\
= & -\nabla\frac{p_{S}}{\rho_A} +  s_{S} \hat{\vec z}.
\label{pres_buoy}
\end{align}

%
%

\subsection{Temperature diffusion}

By invoking equation (\ref{thermodynamics_appendix}), the temperature diffusion term results in
\begin{align}
\label{tempdiff}
\nabla^{2}T_{S} =  \nabla^2\left(T_As_{S}+D\frac{p_{S}}{\rho_A}\right)
\end{align}

\subsection{Governing equations}

By applying equations (\ref{pres_buoy}) and  (\ref{tempdiff}) to (\ref{continuity2}-\ref{momentum2}), (\ref{state2}), (\ref{energy2_entropy}) and (\ref{thermodynamics_appendix}) in the limit $\epsilon=0$ the anelastic equations in entropy formulation appear,
\begin{align}
& \nabla\cdot(\rho_A\vec v)=0
\label{anelasticcontinuity} \\
& \partial_{t}\vec v + (\vec v\cdot \nabla)\vec v = - \nabla\frac{p_{S}}{\rho_A} + s_{S} \hat{\vec z} + \sqrt{\frac{Pr}{Ra}} \left[\nabla^{2}\vec v + \frac{1}{3}(\nabla\cdot\vec v)\right]
\label{anelasticmomentum} \\
& \rho_AT_A\left[\partial_{t}s_{S} (\vec v\cdot\nabla) s_{S}\right]= \frac{1}{\sqrt{Ra Pr}}\nabla^2\left(T_As_{S}+D\frac{p_{S}}{\rho_A}\right) + 2D\sqrt{\frac{Pr}{Ra}}\left[e_{ij} - \frac{1}{3}(\nabla\cdot\vec v)\delta_{ij}\right]^{2}.
\label{anelasticenergy}
\end{align}

\noindent The constant temperature boundary conditions can be implemented by using the linearized form of the thermodynamic relation for entropy as given by equation (\ref{thermodynamics_appendix}). The entropy formulation of the anelastic approximation (\ref{anelasticcontinuity}-\ref{anelasticenergy}) can also be derived directly from the anelastic equations in temperature formulation (\ref{continuity3}-\ref{state3}) by using (\ref{thermodynamics_appendix}). Both formulations are fully equivalent. Note that the the temperature diffusion term is often neglected and replaced by a parametrized entropy-based large eddy diffusion model (e.g. \citealp{Glatzmaier1984,Lantz1999,Jones2011,Gastine2014}).

\section{STEREOSCOPIC 3-D ILLUSTRATIONS OF TURBULENT CONVECTION}

\begin{figure*}[h!]
\setlength{\unitlength}{\linewidth}
\begin{picture}(1.0,1.05)
\put(0.1,0.525){\includegraphics[width=0.8\linewidth]{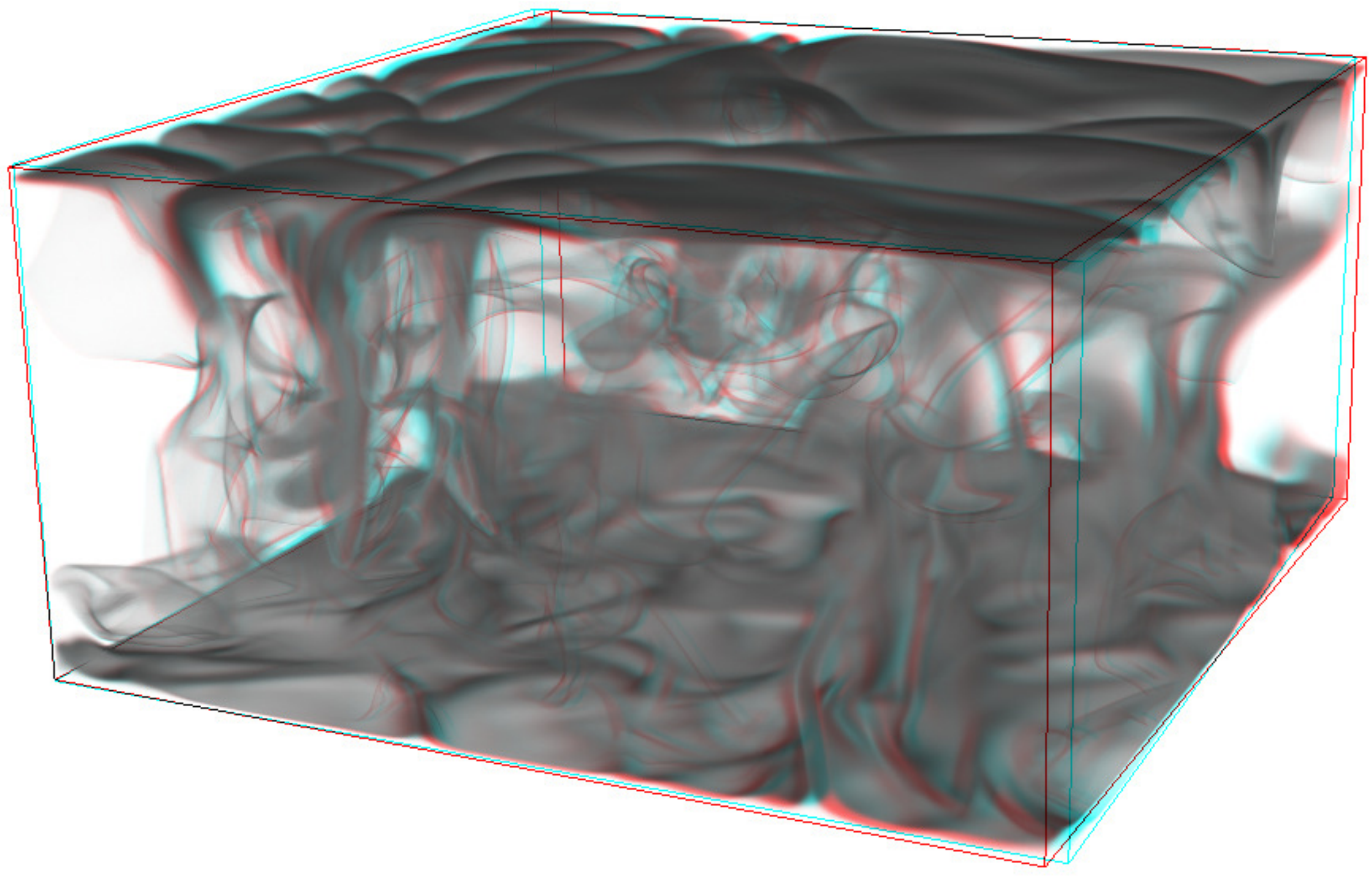}}
\put(0.1,0.0){\includegraphics[width=0.8\linewidth]{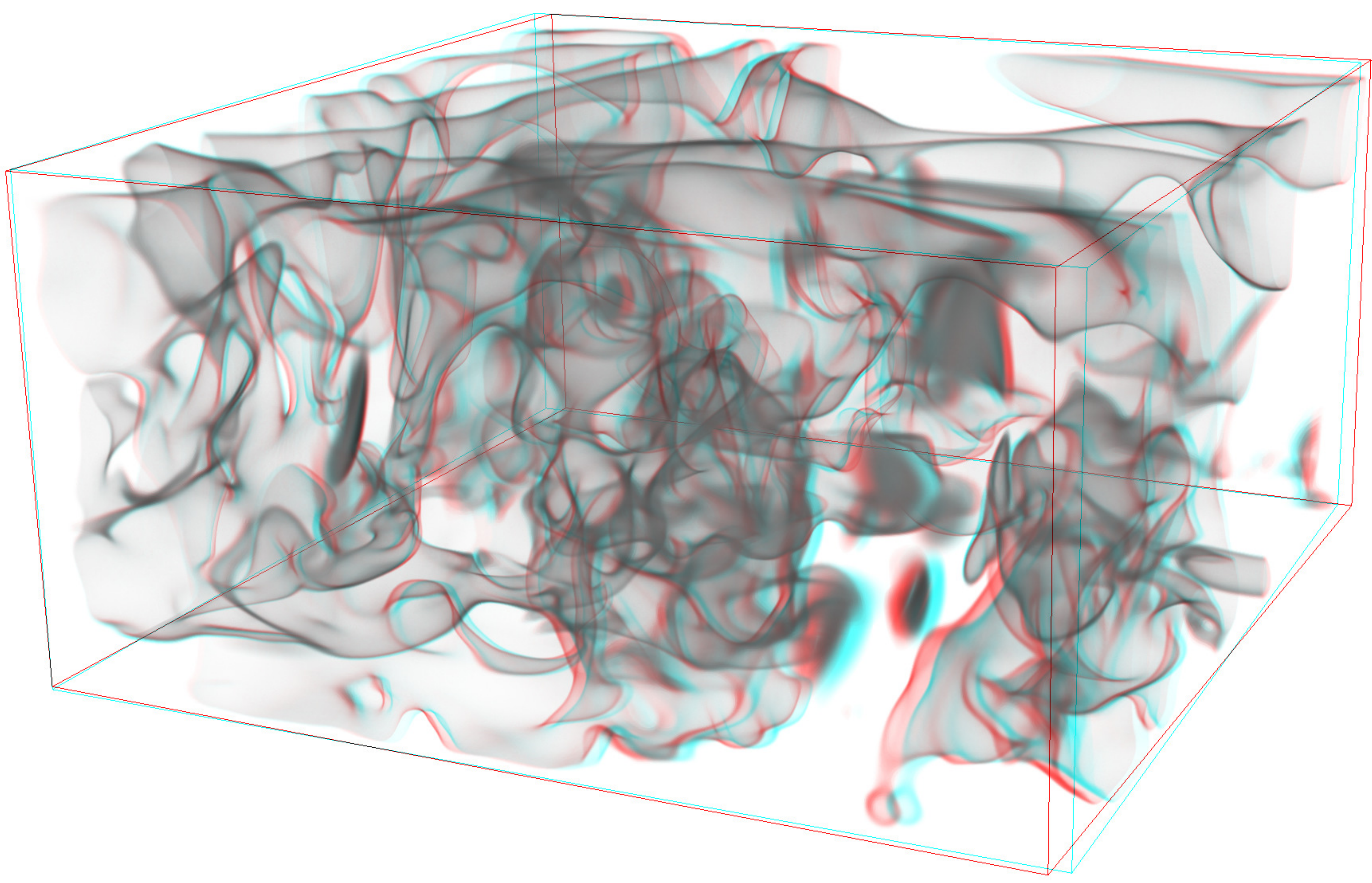}}
\put(0.0,1.025){a)}
\put(0.0,0.5){b)}
\end{picture}
\caption{Anaglyph 3-d version of figure \ref{snapshot_fig} that reflects the full 3-d structures when viewing with red-cyan filter glasses.}
\label{anaglyph}
\end{figure*}

Typical volume renderings of the superadiabatic temperature $T_S$ (a) and vertical velocity $v_z$ (b) for an anelastic simulation run that reached statistical equilibrium are shown in figure \ref{anaglyph}. This stereoscopic 3-d version of figure \ref{snapshot_fig} reflects the full 3-d structures when wearing red-cyan filter glasses.

\bibliographystyle{natbib}
\bibliography{Literature}

\clearpage

\end{document}